\documentclass[12pt,preprint]{aastex}
\usepackage{epsfig} \usepackage{graphicx} \usepackage{amssymb}
\newcommand{\WMAP}{{\sl WMAP}}

\newcommand{\lap}{\la}

\newcommand{\run}{$dn_s/d\ln k$} 
\newcommand{\be}{\begin{equation}}
\newcommand{\ee}{\end{equation}} 
\newcommand{\sh}{{\scriptscriptstyle H}}
\newcommand{\sv}{{\scriptscriptstyle V}} 
\newcommand{\chieff}{{\chi^2_{eff}}}
\newcommand{\map}{{\sl WMAP\ }} 
\newcommand{\cmb}{{\sl WMAP}ext} 
\newcommand{\mpl}{M_{\rm pl}}

\shorttitle{IMPLICATIONS  FOR INFLATION} \shortauthors{PEIRIS ET AL.}
\begin{document}
\title{ FIRST YEAR {\sl WILKINSON MICROWAVE ANISOTROPY PROBE} ({\sl
 WMAP}\altaffilmark{1}) OBSERVATIONS: IMPLICATIONS  FOR INFLATION }%
\author{%
H. V. Peiris\altaffilmark{2}, 
E. Komatsu\altaffilmark{2},  
L. Verde\altaffilmark{2,3}, 
D. N. Spergel\altaffilmark{2}, 
C. L. Bennett\altaffilmark{4}, 
M. Halpern\altaffilmark{5}, 
G. Hinshaw \altaffilmark{4}, 
N. Jarosik\altaffilmark{6}, 
A. Kogut\altaffilmark{4}, 
M. Limon\altaffilmark{4,7}, 
S. S. Meyer\altaffilmark{8}, 
L. Page\altaffilmark{6}, 
G. S. Tucker\altaffilmark{4,7,9}, 
E. Wollack\altaffilmark{4}, 
E. L. Wright\altaffilmark{10} }

\altaffiltext{1}{\map is the result of a partnership between Princeton
                 University and NASA's Goddard Space Flight
                 Center. Scientific guidance is provided by the \map\
                 Science Team.}  
\altaffiltext{2}{Dept of Astrophysical Sciences, Princeton University, Princeton, NJ 08544}
\altaffiltext{3}{Chandra Fellow}
\altaffiltext{4}{Code 685, Goddard Space Flight Center, Greenbelt, MD 20771} 
\altaffiltext{5}{Dept. of Physics and Astronomy, University of British Columbia, Vancouver, BC  Canada V6T 1Z1}
\altaffiltext{6}{Dept. of Physics, Jadwin Hall, Princeton, NJ 08544}
\altaffiltext{7}{National Research Council (NRC) Fellow}
\altaffiltext{8}{Depts. of Astrophysics and Physics, EFI and CfCP, University of Chicago, Chicago, IL 60637} 
\altaffiltext{9}{Dept. of Physics, Brown University, Providence, RI 02912}
\altaffiltext{10}{UCLA Astronomy, PO Box 951562, Los Angeles, CA 90095-1562} 
\email{hiranya@astro.princeton.edu}
\begin{abstract}
 We confront predictions of inflationary scenarios with the  \map
 data, in combination with complementary small-scale CMB  measurements
 and large-scale structure data. The \map detection  of a large-angle
 anti-correlation in the temperature--polarization  cross-power
 spectrum is the signature of adiabatic superhorizon  fluctuations at
 the time of decoupling. The \map data are  described by pure
 adiabatic fluctuations: we place an upper  limit on a correlated CDM
 isocurvature component. Using \map  constraints on the shape of the
 scalar power spectrum  and the amplitude of gravity waves, we
 explore the  parameter space of inflationary models that is
 consistent with  the data. We place limits on inflationary models;
 for example, a minimally-coupled $\lambda\phi^4$ is disfavored at
 more than  3-$\sigma$ using \map data in combination with
 smaller scale CMB and large scale structure survey data. The limits
 on the primordial parameters using \map data alone are: $n_s(k_0=0.002\ {\rm
 Mpc}^{-1})=1.20_{-0.11}^{+0.12}$, \run  $=\ensuremath{-0.077^{+
 0.050}_{- 0.052}}$, $A(k_0=0.002\ {\rm Mpc}^{-1})=
 0.71^{+0.10}_{-0.11}$ (68\% CL),  and $r(k_0=0.002\ {\rm
 Mpc}^{-1})<1.28$ (95\% CL). 

\end{abstract}
\section{INTRODUCTION} \label{intro}

An epoch of accelerated expansion in the early universe, inflation,
dynamically resolves cosmological puzzles such as homogeneity,
isotropy,  and flatness of the universe
\citep{guth:1981,linde:1982,albrecht/steinhardt:1982,sato:1981}, and
generates superhorizon fluctuations without appealing to  fine-tuned
initial setups
\citep{mukhanov/chibisov:1981,hawking:1982,guth/pi:1982,starobinsky:1982,bardeen/steinhardt/turner:1983,mukhanov/feldman/brandenberger:1992}.
During the accelerated expansion phase, generation and amplification
of quantum fluctuations in scalar fields are unavoidable
\citep{parker:1969,birrell/davies:1982}. These fluctuations become
classical after crossing the event horizon. Later during the
deceleration phase they re-enter the horizon, and  seed the matter and
the radiation fluctuations observed in the universe.

The majority of inflation models predict Gaussian, adiabatic, nearly
scale-invariant primordial fluctuations. These properties are generic
predictions of inflationary models. The cosmic microwave
background (CMB) radiation anisotropy  is a promising tool for testing
these properties, as the linearity of the CMB anisotropy preserves basic
properties of the primordial fluctuations. In companion papers,
\citet{spergel/etal:2003} find that adiabatic  scale-invariant
primordial fluctuations fit the \map CMB data as well as  a host of
other astronomical data sets including the galaxy and the
Lyman-$\alpha$ power spectra; \citet{komatsu/etal:2003} find that the
\map CMB data is consistent with Gaussian primordial fluctuations.
These results indicate that predictions of the most basic
inflationary models are in good agreement with the data.

While the inflation paradigm has been very successful, radically
different inflationary models yield similar predictions for the
properties of fluctuations: Gaussianity, adiabaticity, and
near-scale-invariance. To break the degeneracy among the models, we
need to measure the primordial fluctuations precisely.  Even a slight
deviation from Gaussian, adiabatic, near-scale-invariant fluctuations
can place strong constraints on the models
\citep{liddle/lyth:CIALSS}. The CMB anisotropy arising from 
primordial gravitational waves can also be a powerful method for model
testing. In this paper, we confront predictions of various inflationary
models with the CMB data from the {\sl WMAP},  CBI
\citep{pearson/etal:2002}, and ACBAR  \citep{kuo/etal:2002}
experiments, as well as the 2dFGRS \citep{percival/etal:2001} and
Lyman-$\alpha$  power spectra \citep{croft/etal:2002,
gnedin/hamilton:2002}.  

This paper is organized as follows.  In \S~\ref{paradigm}, we show
that the \map detection of an anti-correlation between the temperature
and the polarization fluctuations at $l \sim 150$ is the
distinctive signature of adiabatic superhorizon fluctuations. We
compare the data with specific predictions of inflationary models:
single-field models in \S~\ref{singlemodels}, and double-field models
in \S~\ref{multimodels}. We examine the evidence for features in
the inflaton potential in \S~\ref{features}. Finally, we summarize
our results and draw conclusions in \S~\ref{finish}.

\section{IMPLICATIONS OF \map ``TE'' DETECTION FOR THE INFLATIONARY PARADIGM}
\label{paradigm}

A fundamental feature of inflationary models is a period of
accelerated expansion in the very early universe. During this time,
quantum fluctuations are highly amplified, and their wavelengths are
stretched to outside the Hubble horizon. Thus, the generation of
large-scale fluctuations is an inevitable feature of inflation. These
fluctuations are coherent on what appear to be superhorizon scales at
decoupling.  Without accelerated expansion, the  causal horizon at
decoupling is $\sim 2$ degrees. Causality implies that the correlation
length scale for fluctuations can be no larger than this scale.  Thus,
the detection of superhorizon fluctuations is a distinctive signature
of this early epoch of acceleration.

The {\sl COBE} DMR detection of large scale fluctuations has been
sometimes described as a detection of superhorizon scale fluctuations.
While this is the most likely interpretation of the {\sl
COBE} results, it is not unique.  There are several possible
mechanisms for generating large-scale temperature fluctuations.  For
example, texture models predict a nearly scale-invariant
spectrum of temperature fluctuations on large angular scales
\citep{pen/spergel/turok:1994}. The {\sl COBE} detection sounded the
death knell for these particular models not through its detection of
fluctuations, but due to the low amplitude of the observed fluctuations.
The detection of acoustic temperature fluctuations is also sometimes
evoked as the definitive signature of superhorizon scale fluctuations
\citep{hu/white:1997}. String and defect models do not produce sharp
acoustic peaks
\citep{albrecht/etal:1996,turok/pen/seljak:1998}. However, the
detection of acoustic peaks in the temperature angular power spectrum
does not prove that the fluctuations are superhorizon, as causal
sources acting purely through gravity can exactly mimic the
observed peak pattern \citep{turok:1996b,turok:1996}.  The recent
study of causal seed models by \citet{durrer/kunz/melchiorri:2002}
shows that they can reproduce much of the observed peak structure and
provide a plausible fit to the pre-\map CMB data.

The large-angle $(50 \lap l \lap 150)$ temperature-polarization
anti-correlation detected by \map \citep{kogut/etal:2003} is a
distinctive signature of superhorizon adiabatic fluctuations
\citep{spergel/zaldarriaga:1997}.  The reason for this conclusion is
explained as follows.  Throughout this section, we consider only
scales larger than the sound horizon at the decoupling epoch.
\cite{zaldarriaga/harari:1995} show that, in the tight coupling
approximation, the polarization signal arises from the gradient of the
peculiar velocity of the photon fluid, $\Theta_1$,
\begin{equation}
 \Delta_E \simeq -0.17(1-\mu^2)\Delta\eta_{dec} k\Theta_1(\eta_{dec}),
  \label{eq:polarization}
\end{equation}
where $\Delta_E$ is the $E$-mode (parity-even) polarization
fluctuation, $\eta_{dec}$ is the conformal time at decoupling,
$\Delta\eta_{dec}$ is the thickness of the surface of last scattering in
conformal time, and $\mu=\cos(\hat{k}\cdot\hat{n})$.  The velocity
gradient generates a quadrupole temperature anisotropy  pattern around
electrons which, in turn, produces the $E$-mode polarization.  Note
that while reionization violates the assumptions of tight coupling,
the existence of clear acoustic oscillations in the
temperature-polarization (TE) and temperature-temperature (TT) angular
power spectra imply that most  ($\sim 85\%$) CMB photons detected by
\map did indeed come from $z = 1089$ where the tight coupling
approximation is valid.  The velocity $\Theta_1$ is related to the
photon density fluctuations, $\Theta_0$, through the continuity
equation,   $k\Theta_1 = -3 \left( \dot \Theta_0 + \dot \Phi \right)$,
where $\Phi$ is Bardeen's curvature perturbation.  The observable
temperature fluctuations on large scales are approximately given by
$\Delta_T=\Theta_0(\eta_{dec})+\Psi(\eta_{dec})$, where $\Psi$ is the
Newtonian potential, which equals $-\Phi$ in the absence of 
anisotropic stress.  Therefore, roughly speaking, the photon density
fluctuations generate temperature fluctuations, while the velocity
gradient generates  polarization fluctuations.

The tight coupling approximation implies that the baryon photon fluid
is governed by a single second-order differential equation which
yields a series of acoustic peaks  \citep{peebles/yu:1970,
hu/sugiyama:1995}:
\begin{equation}
 { (\ddot \Theta_0 +\ddot \Phi)}+  \frac{\dot a}{a}\frac{R}{1+R} (\dot
  \Theta_0 + \dot \Phi) +k^2c_s^2 (\Theta_0 +\Phi) = k^2\left(c_s^2
  \Phi -\frac{\Psi}{3}\right),
\end{equation}
where the sound speed $c_s$ is given by
$c_s^2=\left[3(1+R)\right]^{-1}$.  The large-scale solution to this
equation is \citep{hu/sugiyama:1995}
\begin{equation}
 \Theta_0(\eta)+\Phi(\eta)  = \left[\Theta_0(0)+\Phi(0)\right]
  \cos(kc_s\eta) + kc_s\int_0^\eta d\eta'\left[\Phi(\eta')-
  \Psi(\eta')\right]  \sin[k c_s(\eta - \eta')],
\label{eq:dens}
\end{equation}
and the continuity equation gives the solution for the peculiar velocity,
\begin{equation}
 \frac1{3c_s}\Theta_1(\eta)  = \left[\Theta_0(0)+\Phi(0)\right] \sin(k
  c_s\eta)  -kc_s\int_0^\eta
  d\eta'\left[\Phi(\eta')-\Psi(\eta')\right]  \cos[k c_s(\eta -
  \eta')].
\label{eq:vel}
\end{equation}
These solutions (equations (\ref{eq:polarization}), (\ref{eq:dens}),
and (\ref{eq:vel}))  are valid regardless of the nature of the source
of fluctuations.

In inflationary models, a period of accelerated expansion generates
superhorizon adiabatic fluctuations, so that the first term in
equation (\ref{eq:dens}) and (\ref{eq:vel}) is non-zero.  Since
$\Psi\simeq-\Phi$ and
$\Theta_0(0)+\Phi(0)=\frac32\Phi(0)=\frac53\Phi(\eta_{dec})$ on
superhorizon scales, one obtains $\Delta_T \simeq
-\frac13\Phi(\eta_{dec})\cos(kc_s\eta_{dec})$, and  $\Delta_E\simeq
0.17(1-\mu^2)kc_s\Delta\eta_{dec}\Phi(\eta_{dec})\sin(kc_s\eta_{dec})$
(see \citet{hu/sugiyama:1995} and \citet{zaldarriaga/harari:1995} for
derivation).  Therefore, the cross correlation is found to be
\begin{equation}
 \langle\Delta_T\Delta_E\rangle \simeq
  -0.03(1-\mu^2)(kc_s\Delta\eta_{dec})P_\Phi(k)\sin(2kc_s\eta_{dec}),
\end{equation}
where $P_\Phi(k)$ is the power spectrum of $\Phi(\eta_{dec})$.  The observable
correlation function is estimated as
$k^3\langle\Delta_T\Delta_E\rangle$.  Clearly, there is an
anti-correlation peak near $k c_s\eta_{dec}\sim 3\pi/4$, which corresponds
to $l \sim 150$: this is the distinctive signature of primordial
adiabatic fluctuations.  In other words, the anti-correlation appears
on superhorizon scales at decoupling, because of the modulation
between the  density mode, $\cos(kc_s\eta_{dec})$, and the velocity
mode,  $\sin(kc_s\eta_{dec})$, yielding $\sin(2kc_s\eta_{dec})$, which
has a peak on scales larger than the horizon size, $c\eta_{dec}\simeq
\sqrt{3}c_s\eta_{dec}$.

Cosmic strings and textures are examples of active models. In these
models, causal field dynamics continuously generate spatial
variations  in the energy density of a field.
\citet{magueijo/etal:1996} describe the general dynamics of active
models.  These models do not have the first term in
equation (\ref{eq:dens}) and (\ref{eq:vel}), but the fluctuations are
produced by the  second term, the growth of $\Phi$ and $\Psi$.  The
same applies to primordial isocurvature fluctuations, where the
non-adiabatic pressure causes $\Phi$ and $\Psi$ to grow.  While the
problem is more complicated, these  models give a positive correlation
between temperature and  polarization fluctuations on large scales.
This positive correlation is predicted not just for texture
\citep{seljak/pen/turok:1997} and scaling seed models
\citep{durrer/kunz/melchiorri:2002}, but is the generic signature of
{\it any} causal models \citep{hu/white:1997}\footnote{
\cite{hu/white:1997} use an opposite sign convention for the TE  cross
power spectrum.} that lack a period of accelerated expansion.

\begin{figure}
\plotone{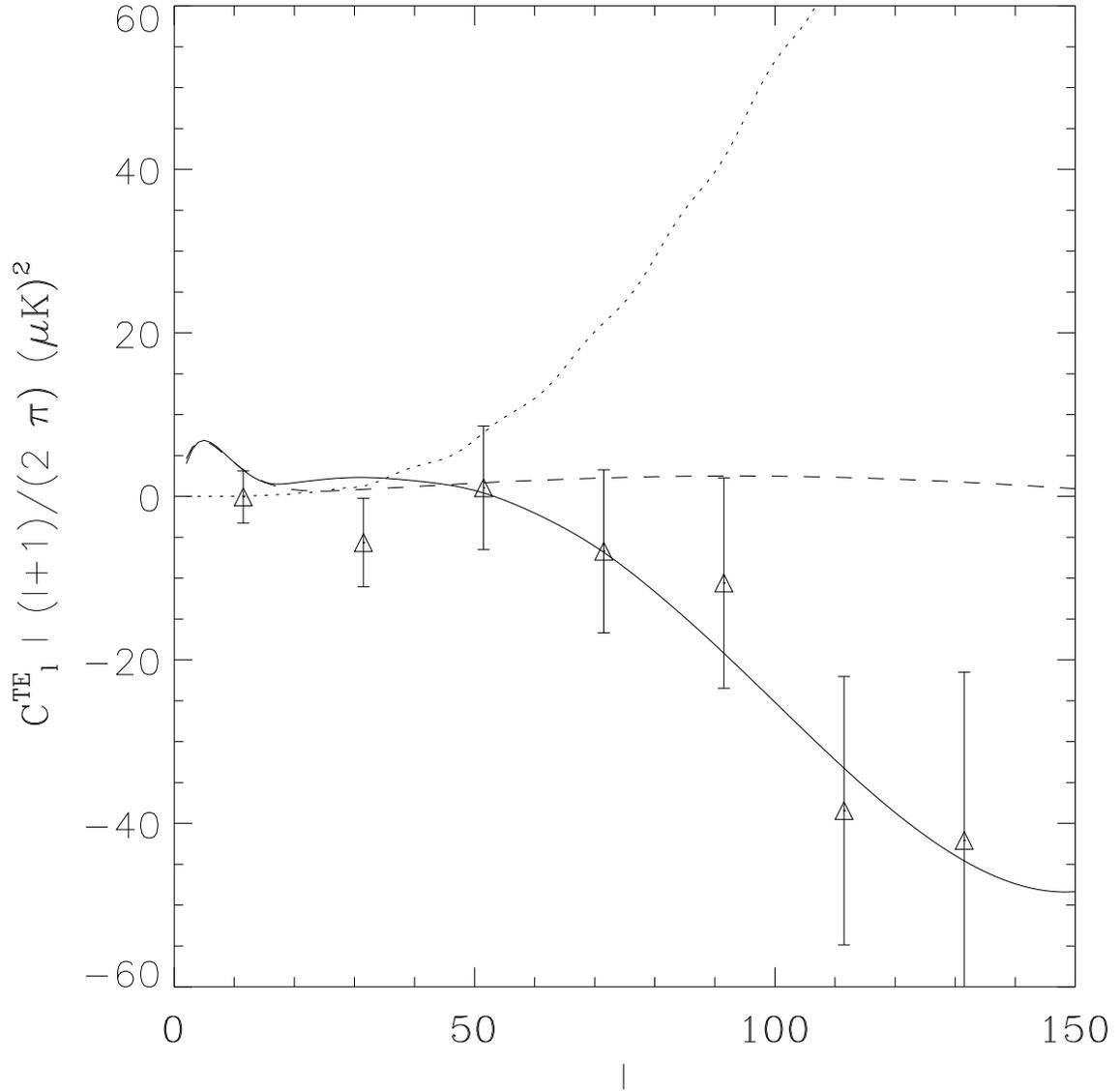}
 \caption{Temperature-Polarization angular power spectrum. The
 large-angle TE power spectrum predicted in primordial adiabatic
 models (solid), primordial isocurvature models (dashed), and in
 causal scaling seed models (dotted). The \map TE data
 \citep{kogut/etal:2003} is shown for comparison, in bins of $\Delta 
 l=10$.}%
\label{fig:tecomp}
\end{figure}

Figure \ref{fig:tecomp} shows the predictions of the TE large angle
correlation predicted in typical primordial adiabatic, isocurvature,
and causal scaling seed models compared with the \map data. The causal
scaling seed model shown is a flat Family I model in the
classification of \citet{durrer/kunz/melchiorri:2002} that provided a
good fit to the pre-\map temperature data. 

The \map detection of a TE anti-correlation at $l\sim 50-150$, scales
that correspond to superhorizon scales at the epoch of decoupling,
rules out a broad class of active models. It implies the existence of
superhorizon, adiabatic fluctuations at decoupling. If these
fluctuations were generated dynamically rather than by setting special
initial conditions then the TE detection requires that the universe
had a period of accelerated expansion. In addition to inflation, the
pre-Big-Bang scenario \citep{gasperini/veneziano:1993} and the
Ekpyrotic scenario \citep{khoury/etal:2001,khoury/etal:2002} predict
the existence of superhorizon fluctuations.

\section{SINGLE FIELD INFLATION MODELS}
\label{singlemodels}

In this section we explore how predictions of specific models that
implement inflation (see \citet{lyth/riotto:1999} for a survey)
compare with current observations.

\subsection{Introduction}

The definition of ``single-field inflation'' encompasses the class of
models in which the inflationary epoch is described by a single scalar
field, the inflaton field.  We also include a class of models called
``hybrid'' inflation models as single-field models. While hybrid
inflation requires a second field to end inflation \citep{linde:1994},
the second field does not contribute to the dynamics  of inflation or
the observed fluctuations.  Thus, the predictions of hybrid inflation
models can be  studied in the context of single-field models.

During inflation the potential energy of the inflaton field $V$
dominates over the kinetic energy. The Friedmann equation then tells
us that the expansion rate, $H$, is nearly constant in time: $H\equiv
\dot{a}/a\simeq \mpl^{-1} (V/3)^{1/2}$, where $\mpl\equiv (8\pi
G)^{-1/2} = m_{\rm pl}/\sqrt{8\pi}= 2.4\times 10^{18}~{\rm GeV}$ is
the reduced Planck energy.  The universe thus undergoes an accelerated
expansion phase, expanding exponentially as $a(t)\propto \exp(\int
Hdt)\simeq \exp(Ht)$.  One usually uses the $e$-folds remaining at a
given time, $N(t)$, as a measure of  how much the universe expands
from $t$ to the end of inflation, $t_{\rm end}$: $N(t)\equiv
\ln[a(t_{\rm end})]-\ln[a(t)]= \int_t^{t_{\rm end}} H(t) dt$.  It is
known that flatness and homogeneity of the universe require $N(t_{\rm
start})> 50$, where $t_{\rm start}$ is the time at the onset  of
inflation (i.e., the universe needs to be expanded to at least
$e^{50}\simeq 5\times10^{21}$ times larger by $t_{\rm end}$).  The
accelerated expansion of this amount dilutes any initial inhomogeneity
and spatial curvature until they become negligible in the observable
universe today.

\subsection{Framework for data analysis}

\subsubsection{Parameterizing the primordial power spectra}

The power spectrum of the CMB anisotropy is determined by the power
spectra of the curvature and tensor perturbations.  Most inflationary
models predict scalar and tensor power spectra that approximately
follow power laws: $\Delta^2_{\cal R}(k) \equiv
k^3/(2\pi^2)\langle\left|{\cal R}_{\mathbf k}\right|^2\rangle\propto
k^{n_s-1}$ and $\Delta^2_h(k) \equiv 2
k^3/(2\pi^2)\langle\left|h_{+{\mathbf k}}\right|^2 +
\left|h_{\times{\mathbf k}}\right|^2\rangle\propto k^{n_t}$. Here,
${\cal R}$ is the curvature perturbation in the comoving gauge, and
$h_+$ and $h_\times$ are the two polarization states of the primordial
tensor perturbation. The spectral indices $n_s$ and $n_t$ vary slowly
with scale, or not at all. As spectral indices deviate more and more
from scale invariance  (i.e., $n_s=1$ and $n_t=0$), the power-law
approximation usually becomes  less and less accurate. Thus, in
general, one must consider the scale dependent ``running'' of the
spectral indices, \run\ and $d n_t/d\ln k$.  We parameterize these
power spectra by
\begin{eqnarray}
 \label{eq:P_R} \Delta^2_{\cal R}(k)&=& \Delta^2_{\cal R}(k_0)
  \left(\frac{k}{k_0}\right)^{n_s(k_0)-1+\frac{1}{2}(dn_s/d\ln
  k)\ln(k/k_0)}, \\ \label{eq:P_h} \Delta^2_h(k)&=& \Delta^2_h(k_0)
  \left(\frac{k}{k_0}\right)^{n_t(k_0)+\frac{1}{2}(dn_t/d\ln
  k)\ln(k/k_0)},
\end{eqnarray}
where $\Delta^2(k_0)$ is a normalization constant, and $k_0$ is some pivot
wavenumber.  The running, $dn/d\ln k$, is defined by the second
derivative of the  power spectrum, $dn/d\ln k\equiv d^2\Delta^2/d\ln k^2$,
for both the  scalar and the tensor modes, and is independent of $k$.
This parameterization gives the definition of the spectral index,
\be
\label{eq:nsdef}
 n_s(k)-1 \equiv \frac{d \ln \Delta^2_{\cal R}}{d \ln k} =
 n_s(k_0)-1+\frac{dn_s}{d\ln k}\ln\left(\frac{k}{k_0}\right), \ee
for the scalar modes, and
\be
\label{eq:ntdef}
 n_t(k) \equiv \frac{d \ln \Delta^2_h}{d \ln k} =  n_t(k_0)+\frac{dn_t}{d\ln
 k}\ln\left(\frac{k}{k_0}\right), \ee
for the tensor modes.  In addition, we re-parameterize the tensor
power spectrum amplitude, $\Delta^2_h(k_0)$, by the ``tensor/scalar ratio
$r$'', the relative amplitude  of the tensor to  scalar modes,
given by\footnote{This definition of $r$ agrees with the definition of
$T/S$ in the {\sf CAMB} code \citep{lewis/challinor/lasenby:2000}
and $r$ in \citet{leach/etal:2002}. We have modified {\sf CMBFAST}
\citep{seljak/zaldarriaga:1996} accordingly to match the same convention.}
\begin{equation}
 \label{eq:rdef} r \equiv \frac{\Delta^2_h(k_0)}{\Delta^2_{\cal R}(k_0)}.
\end{equation}
The ratio of the tensor quadrupole to the scalar quadrupole, $r_2$, is often
quoted when referring to the tensor/scalar ratio. The relation between
$r_2$ and the definition of the tensor/scalar ratio above is somewhat
cosmology-dependent. For an SCDM universe with no reionization, it is: 
\begin{equation}
 \label{eq:r2def} r_2 = 0.8625\ r.
\end{equation}
For comparison, for the maximum likelihood single field inflation model
for the \cmb+2dFGRS data sets presented in the table notes of Table
\ref{table:single_field} in \S\ref{parameters}, this relation is
$r_2=0.6332\ r$.  

Following notational conventions in \cite{spergel/etal:2003}, we use
$A(k_0)$ for the scalar power spectrum amplitude, where $A(k_0)$ and
$\Delta^2_{\cal R}(k_0)$ are related through
\begin{eqnarray}
 \label{eq:Adef} \Delta^2_{\cal R}(k_0) &=& 800 \pi^2
 \left(\frac{5}{3}\right)^2 \frac{1}{T_{CMB}^2} A(k_0) \\
&\simeq& 2.95\times10^{-9}  A(k_0).
\end{eqnarray}
Here, $T_{CMB}=2.725\times 10^6\ (\mu K)$. This relation is derived
in \citet{verde/etal:2003}. One can use equations (\ref{eq:P_R}),
(\ref{eq:nsdef}), and  (\ref{eq:ntdef}) to evaluate $A$, $n_s$, and
$n_t$ at a different wavenumber from $k_0$, respectively. Hence,
\be
\label{eq:conv_A}
  A(k_1)=A(k_0)
  \left(\frac{k_1}{k_0}\right)^{n_s(k_0)-1+\frac{1}{2}(dn_s/d\ln
  k)\ln(k_1/k_0)}.  
\ee

We have 6 observables ($A$, $r$, $n_s$, $n_t$,
$dn_s/d\ln k$, $dn_t/d\ln k$), each of which  can be compared to
predictions of an inflationary model.

The complementary approach (which we do not investigate in this work)
is to parameterize the primordial power spectrum in a
model-independent way (see, for example,
\citet{wang/spergel/strauss:1999}). These authors anticipated that \map
has the potential ability to reveal deviations from scale-invariance
when combined with large scale structure
data. \citet{mukherjee/wang:2003a, mukherjee/wang:2003b} extend this
approach and use it to put model-independent constraints on the
primordial power spectrum using the pre-\map CMB data.


\subsubsection{Slow roll parameters} \label{psr}

In the context of slow roll inflationary models,  only three ``slow-roll
parameters'', plus the amplitude of the  potential, determine the six
observables  ($A$, $r$, $n_s$, $n_t$, $dn_s/d\ln k$, $dn_t/d\ln
k$). Thus, one can  use the relations among the observables to either
reduce the number  of parameters to four, or cross-check if the slow
roll inflation  paradigm is consistent with the data.  The slow-roll
parameters are defined by \citep{liddle/lyth:1992,liddle/lyth:1993}:
%
%
\begin{eqnarray}
 \label{eq:eps} \epsilon_\sv &\equiv&
  \frac{\mpl^2}{2}\left(\frac{V'}{V}\right)^2, \\  \label{eq:eta}
  \eta_\sv  &\equiv& \mpl^2 \left(\frac{V''}{V}\right), \\
  \label{eq:xi} \xi_\sv  &\equiv& \mpl^4
  \left(\frac{V'V'''}{V^2}\right),
\end{eqnarray}
where prime denotes derivatives with respect to the field
$\phi$. Here, $\epsilon_\sv$ quantifies ``steepness'' of the slope of
the  potential which is positive-definite, $\eta_\sv$ quantifies
``curvature'' of  the potential, and $\xi_\sv$, (which is {\it not}
positive-definite, but is unfortunately often denoted $\xi^2$ in the
literature because it is a second order parameter), quantifies the
third derivative of the potential, or ``jerk''. All parameters must be
smaller than one for inflation to occur. We denote these ``potential
slow roll'' parameters with a subscript $V$ to distinguish them from
the ``Hubble slow roll'' parameters of Appendix~\ref{A:floweq}.
\citet{gratton/etal:2003} discuss the equivalent set of parameters for
the Ekpyrotic scenario.

Parameterization of slow roll models by $\epsilon_\sv$, $\eta_\sv$,
and $\xi_\sv$ avoids relying on specific models, and enables one to
explore a large model space without assuming a specific model.  Each
inflation model predicts the slow-roll parameters, and hence  the
observables.  A standard slow roll analysis gives observable
quantities in terms of the slow roll parameters to first order as
(see \cite{liddle/lyth:CIALSS} for a review),
\begin{eqnarray}
 \label{eq:A} \Delta^2_{\cal R}   &=&  \frac{V/\mpl^4}{24\pi^2\epsilon_\sv},\\
  \label{eq:r} r   &=&  16 \epsilon_\sv, \\  \label{eq:n_s} n_s-1 &=&
  -6\epsilon_\sv + 2\eta_\sv = -\frac{3r}{8} + 2\eta_\sv, \\
  \label{eq:n_t} n_t   &=&  -2\epsilon_\sv = -\frac{r}{8}, \\
  \label{eq:run_s} \frac{dn_s}{d\ln k} &=& 16\epsilon_\sv\eta_\sv -
  24\epsilon_\sv^2 - 2\xi_\sv = r \eta_\sv -  \frac3{32}r^2 -
  2\xi_\sv  = -\frac23\left[(n_s-1)^2-4\eta_\sv^2\right] - 2\xi_\sv, \\
  \label{eq:run_t} \frac{dn_t}{d\ln k} &=& 4\epsilon_\sv\eta_\sv -
  8\epsilon_\sv^2 = \frac{r}{8} \left[(n_s-1)+\frac{r}{8}\right].
\end{eqnarray}
The tensor tilt in inflation is always red, $n_t< 0$.  The equation
$n_t=-r/8$ is known as the consistency relation for single-field
inflation models (it weakens to an inequality for multi-field
inflation models).   We use the relation to reduce the number of
parameters.  While we have also carried out the analysis including
$n_t$ as a parameter, and verified that there is a parameter space
satisfying the consistency relation, including $n_t$ obviously weakens
the constraints on the other observables.    Given that we find $r$ is
consistent with zero (\S~\ref{parameters}), the running tensor index
$dn_t/d\ln k$ is  poorly constrained with our data set; thus,  we
ignore it and constrain our models using the other four observables
($A$, $r$, $n_s$, $dn_s/d\ln k$) as free parameters.

\subsection{Determining the power spectrum parameters}
\label{parameters}

We use a Markov Chain Monte Carlo (MCMC) technique to explore the
likelihood surface.  \citet{verde/etal:2003} describe our
methodology. We use the \map TT \citep{hinshaw/etal:2003} and TE
\citep{kogut/etal:2003} angular power spectra.  To measure the shape
of the spectrum (i.e., $n_s$ and $dn_s/d\ln k$) accurately, we want to
probe the primordial power spectrum over as wide a range of scales as
possible. Therefore, we also  include the CBI
\citep{pearson/etal:2002} and ACBAR \citep{kuo/etal:2002} CMB data,
Lyman $\alpha$ forest data \citep{croft/etal:2002,
gnedin/hamilton:2002}, and the 2dFGRS large-scale structure data
\citep{percival/etal:2001} in our likelihood analysis. We refer to the
combined {\sl WMAP}+CBI+ACBAR data as \cmb.

In total, the single field inflation model is described by an
8-parameter model: 4 parameters for characterizing a
Friedmann-Robertson-Walker universe (baryonic density $\Omega_b h^2$,
matter density $\Omega_m h^2$, Hubble constant in units of $100\ {\rm
kms}^{-1} {\rm Mpc}^{-1}$ $h$, optical depth $\tau$), and 4 parameters
for the primordial power spectra ($A$, $r$, $n_s$, \run). When we add
2dFGRS data, we need two further large-scale structure parameters,
$\beta$ and $\sigma_p$,  to marginalize over the shape and the
amplitude of the 2dFGRS power spectrum \citep{verde/etal:2003}. We run
MCMC with these eight (\WMAP\ only model) or ten (\cmb+2dFGRS,
\cmb+2dFGRS+Ly $\alpha$ models)  parameters in order to get our
constraints.

The priors on the model are: a flat universe, a cosmological constant
equation of state for the dark energy, and a restriction of
$\tau<0.3$.  
\clearpage
\begin{deluxetable}{lccc}
\rotate
\tablecaption{Parameters For Primordial Power Spectra: Single Field
Inflation Model \label{table:single_field}}
\tablewidth{0pt} \tablehead{\colhead{Parameter} &
\colhead{\WMAP\tablenotemark{a}} &
\colhead{\cmb+2dFGRS\tablenotemark{a}} &
\colhead{\cmb+2dFGRS$+$Lyman $\alpha$\tablenotemark{a}} }
\startdata
$n_s(k_0=0.002\ {\rm Mpc}^{-1})$& $1.20_{-0.11}^{+0.12}$ &
$1.18_{-0.11}^{+0.12}$  & $1.13 \pm 0.08$ \\ 
$r(k_0=0.002\ {\rm Mpc}^{-1})$ & $<1.28/0.81/0.47\tablenotemark{b}$ & $<1.14/0.53/0.37\tablenotemark{b}$   & $<0.90/0.43/0.29\tablenotemark{b}$ \\ 
\run  & $\ensuremath{-0.077^{+ 0.050}_{- 0.052}}$ & $\ensuremath{-0.075^{+0.044}_{-0.045}}$ & $\ensuremath{-0.055^{+ 0.028}_{- 0.029}}$  \\ 
$A(k_0=0.002\ {\rm Mpc}^{-1})$ & $0.71^{+0.10}_{-0.11}$ & $0.73 \pm 0.09$ & $0.75^{+0.08}_{-0.09}$ \\ 
$\Omega_bh^2$ & $0.024 \pm 0.002$ & $0.023 \pm 0.001$ & $0.024 \pm 0.001$ \\
$\Omega_mh^2$ & $0.127 \pm 0.017$ & $0.134 \pm 0.006$ & $0.134 \pm 0.006$ \\
$h$	      & $0.78 \pm 0.07$ & $0.75^{+0.03}_{-0.04}$  & $0.75 \pm 0.03$ \\
$\tau$	      & $0.22 \pm 0.06$ & $0.20 \pm 0.06$ & $0.18 \pm 0.06$ \\
$\sigma_8$    & $0.82^{+0.13}_{-0.12}$ & $0.85 \pm 0.05$ & $0.85 \pm 0.05$

\enddata \tablenotetext{a}{The quoted values are the mean and the 68\%
probability level of the 1--d marginalized likelihood. For both
\cmb+2dFGRS and \cmb+2dFGRS$+$Lyman $\alpha$ data sets, the 10--d
maximum likelihood point in the Markov Chain ($1.5\times10^6$ steps) for this
model is [$\Omega_b h^2=0.024$, $\Omega_m h^2=0.132$, $h=0.77$,
$n(k_{0.002})=1.15$, $r(k_{0.002})=0.42$, $dn_s/d\ln k=-0.052$,
$A(k_{0.002})=0.75$, $\tau=0.21$, $\sigma_8=0.87$]. Here, 
$k_{0.002}$ is $k_0=0.002\ {\rm Mpc}^{-1}$. The maximum likelihood
model in the MCMC using \WMAP\ data alone is [$\Omega_b h^2=0.023$, $\Omega_m h^2=0.122$, $h=0.79$,
$n(k_{0.002})=1.27$, $r(k_{0.002})=0.56$, $dn_s/d\ln k=-0.10$,
$A(k_{0.002})=0.74$, $\tau=0.29$]. Great care must be taken in
interpreting this point. It is given here for completeness
only, and we do not recommend it for use in any analysis. There is a
long, flat degeneracy between $n$ and $\tau$, as described in \S3
\citet{spergel/etal:2003}, and this point happened to lie at the very
blue edge of this degeneracy right at the edge of our upper limit prior on
$\tau$. This Markov chain had extra freedom because we are adding
three parameters over the model discussed in \citet{spergel/etal:2003},
thereby introducing significant new degeneracies (see Figure
\ref{fig:obsspace}).}  
\tablenotetext{b}{The 95\% upper limits for the tensor-scalar ratio are
quoted for various priors in the following order: [no prior on \run\
or $n_s$] / [\run$=0$] / [$n_s<1$]. The priors were applied to the
output of the MCMC.} 
\end{deluxetable}
\clearpage
Table \ref{table:single_field} shows results of our analysis for the
\WMAP, \cmb + 2dFGRS and \cmb + 2dFGRS + Lyman $\alpha$ data sets. We
evaluate $n_s$, $A$, and $r$ in the fit at $k_0=0.002\
\mathrm{Mpc^{-1}}$. Thus, this table and the figures to follow report
the results for $A$ and $n_s$ at $k_0=0.002\ \mathrm{Mpc^{-1}}$ . Note
that \citet{spergel/etal:2003} report these quantities evaluated at
$k_0=0.05\ \mathrm{Mpc^{-1}}$ (using equations (\ref{eq:conv_A}) and
(\ref{eq:nsdef})). There are $3.2$ $e$--folds between $k_0=0.002\
\mathrm{Mpc^{-1}}$ and $k_0=0.05\ \mathrm{Mpc^{-1}}$.

We did not find any tensor modes. Table \ref{table:single_field} shows
95\% upper limits for the tensor-scalar ratio $r$ at $k=0.002\ {\rm
Mpc}^{-1}$, for various combinations of the data sets. As we will see
later, there are strong degeneracies present between the parameters
$n_s$, $r$ and $\ensuremath{dn_s/d\ln{k}}$. For example, one can add
power at low multipoles by increasing $r$ and then remove it with a
bluer $n_s$ while keeping the low $l$ amplitude constant. Thus, one
can obtain stronger constraints on $r$ by assuming different priors on
$n_s$ and $\ensuremath{dn_s/d\ln{k}}$. In the table, we list the 95\% CL
constraints on $r$ that would be obtained if (1) there were no priors
on $n_s$ or $\ensuremath{dn_s/d\ln{k}}$, (2) if one only considers
models with no running of the scalar spectral index, and (3) if only
models with red spectral indices are considered (non-hybrid-inflation
models predict red indices in general). 

The no-prior $r$ limit $r<0.9$, along with the 2--$\sigma$ upper limit
on the amplitude $A(k=0.002\ {\rm Mpc}^{-1})<0.75+0.08\times2$,
implies that the energy scale of inflation $V^{1/4} < 3.3 \times
10^{16}$ GeV at the 95\% confidence level.    

Note that in the case of the \WMAP-only Markov chain, the degeneracy
between $n_s$, $r$ and $\ensuremath{dn_s/d\ln{k}}$ is cut off by the
prior $\tau<0.3$ ($\tau$ is denegerate with $n_s$). Thus, a better
upper limit on $\tau$ will significantly tighten the constraints on
this model from the \WMAP\ data alone.

All cosmological parameters are consistent with the best-fit running
model of \citet{spergel/etal:2003}, which was obtained for a
$\Lambda$CDM model with no tensors and a running spectral index.
Adding the extra parameter $r$ does not improve the fit.

Our constraint on $n_s$ shows that the scalar power spectrum is nearly
scale-invariant.  One implication of this result is that fluctuations
were generated during accelerated expansion in nearly de-Sitter space
\citep{mukhanov/chibisov:1981,hawking:1982,guth/pi:1982,starobinsky:1982,bardeen/steinhardt/turner:1983,mukhanov/feldman/brandenberger:1992},
where the equation of state of the scalar field is $w\simeq -1$.
Recently, \citet{gratton/etal:2003} have shown that there is only one
other possibility for robustly obtaining adiabatic fluctuations with
nearly scale-invariant spectra: $w\gg 1$. The Ekpyrotic/Cyclic
scenarios correspond to this case. Note, however, that predictions for
the primordial perturbation spectrum resulting from the Ekpyrotic
scenario are controversial (see, for example,
\citet{tsujikawa/brandenberger/finelli:2002}).

We find a marginal $2 \sigma$ preference for a running spectral index
in all three data sets; \ensuremath{dn_s/d\ln{k} = -0.055^{+ 0.028}_{-
0.029}} (\cmb+2dFGRS+Lyman $\alpha$ data set). This same preference
was seen in the analysis without tensors carried out in
\citet{spergel/etal:2003}.

\begin{figure}
\plotone{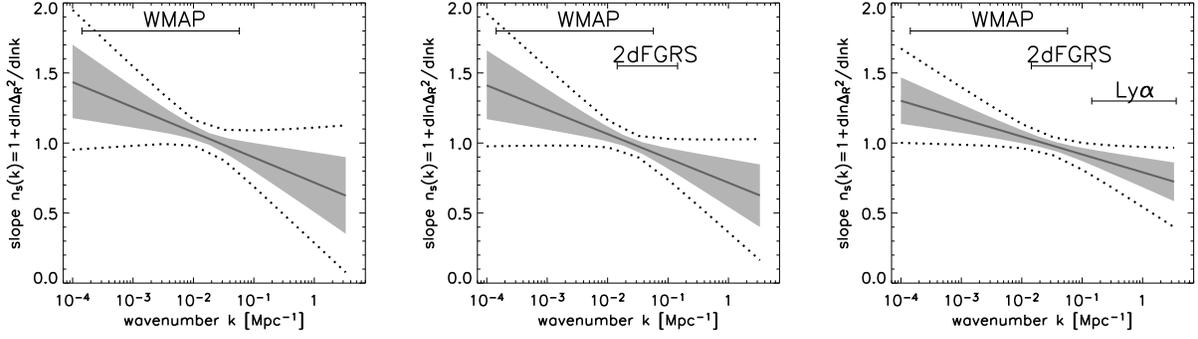}  
\caption{ This figure shows $n_s$ as a function of
 $k$ for the \WMAP\ (left), \cmb+2dFGRS (middle) and
 \cmb+2dFGRS$+$Lyman $\alpha$ (right) data sets. The mean (solid line)
 and the 68\% (shaded area) and 95\% (dashed lines) intervals are
 shown. The scales probed by {\sl WMAP}, 2dFGRS and Lyman $\alpha\ $
 are indicated on the figure.  }%
\label{fig:nkcmb2dflya}
\end{figure}

Figure~\ref{fig:nkcmb2dflya} shows our constraint on $n_s$ as a
function of $k$ for the \WMAP, \cmb + 2dFGRS and \cmb + 2dFGRS + Lyman
$\alpha$ data sets. At each wavenumber $k$, we use equation
(\ref{eq:nsdef}) to convert $n_s(k_0=0.002~{\rm Mpc}^{-1})$ to
$n_s(k)$ at each wavenumber.  Then, we evaluate the mean (solid line),
68\% interval (shaded area),  and 95\% interval (dashed lines) from
the MCMCs. This shows a hint that the spectral index is running from
blue ($n_s>1$) on large scales to red ($n_s<1$) on small scales. In
our MCMCs, for the \WMAP\ data set alone, 91\% of models explored by the
chain have a scalar spectral index running from blue at $k=0.0007\
\mathrm{Mpc^{-1}}$ ($l\sim10$) to red at $k=2\ \mathrm{Mpc^{-1}}$. For
the \cmb+2dFGRS data set, 95\% of models go from a blue index at large
scales to a red index at small scales, and when Lyman $\alpha$ forest
data is added, the fraction running from blue to red becomes 96\%.


One-loop correction and renormalization usually predict running mass
and/or running coupling constant, giving some \run. Detection of it
implies interesting quantum phenomena during inflation (see
\citet{lyth/riotto:1999} for a review). For the running of the scalar
spectral index (equation \ref{eq:run_s}), 
\begin{equation}
 \label{eq:run_s*}
  \frac{dn_s}{d\ln k} = -2\xi_\sv - \frac23\left[(n_s-1)^2-4\eta_\sv^2\right]. 
\end{equation}
Since the data requires $n_s \sim 1$ (see Table
\ref{table:single_field}), $(n_s-1)^2 \lap 0.01$.  It is especially
small when $n_s-1\simeq 2\eta_\sv$, (see Case A and Case D in
\S~\ref{hsr}).  Therefore, if $dn_s/d\ln k$ is large enough to detect,
$dn_s/d\ln k > 10^{-2}$, then $dn_s/d\ln k$ must be dominated by
$2\xi_\sv$, a product of the first and the third derivatives of the
potential (equation (\ref{eq:xi})).  The hint of $dn_s/d\ln k$ in our
data can be interpreted as  $\xi_\sv\simeq -\frac{1}{2} dn_s/d\ln k=
0.028\pm 0.015$.  However, obtaining the running from blue to red,
which is suggested by the data, may require fine-tuned properties in
the shape of the potential. More data are required to determine
whether the hints of a running index are real.

\subsection{Single field models confront the data}

\subsubsection{Testing a specific inflation model: $\lambda\phi^4$}
\label{phi4}

As a prelude to showing constraints on broad classes of
inflationary models, we first illustrate the power of the data using
the example of the minimally-coupled $V=\lambda \phi^4/4$ model, 
which is often used as an introduction to inflationary models
\citep{linde:1990}. We show that this textbook example is unlikely.
%
%

The Friedmann and continuity equations for a homogeneous scalar field
lead to the slow-roll parameters, which one can use in conjunction
with the equations of \S~\ref{psr} in order to obtain predictions for
the observables.  
For the potential $V(\phi)=\lambda \phi^4 /4$, one obtains the
potential slow roll parameters as:
\begin{equation}
 \epsilon_\sv =8 \frac{\mpl^2}{\phi^2},\qquad
 \eta_\sv = {1}{2}\frac{\mpl^2}{\phi^2},\qquad
 \xi_\sv = 96\frac{\mpl^4}{\phi^4}.
\end{equation}
The number of $e$-foldings remaining till the end of inflation is defined by
\be
N = \int_t^{t_{end}} H dt \simeq
\frac{1}{\mpl^2} \int_{\phi_{end}}^{\phi} \frac{V}{V'}\ d\phi =
\frac18\left(\frac{\phi^2 - \phi_{end}^2}{\mpl^2}\right), 
\ee
where $\epsilon_\sv(\phi_{end})=1$ defines the end of
inflation. Assuming $\phi_{end}\ll\phi$, taking the horizon exit scale
as $\phi \simeq \sqrt{8N}\mpl$ and $N=50$, one obtains $n_s=0.94$ and
$r=0.32$ using equations (\ref{eq:r}) and (\ref{eq:n_s}). As \run\ is
negligible for this model, we use  $dn_s/d\ln k=0$.

We maximize the likelihood for this model by running  a simulated
annealing code.  We fit to \cmb+2dFGRS data, varying the following
parameters: $\Omega_b h^2$, $\Omega_m h^2$, $h$, $\tau$,
$A$\footnote{While $A$ is an inflationary parameter, it is directly
related to the self-coupling $\lambda$ which we  do not know; thus, we
treat it as a parameter.}, $\beta$, and $\sigma_p$, while keeping
$n_s$, \run, and $r$ fixed at the $\lambda\phi^4$ values.  The maximum
likelihood model obtained has [$\Omega_b h^2=0.022$, $\Omega_m
h^2=0.135$, $\tau=0.07$, $A=0.67$, $h=0.69$, $\sigma_8=0.76$].  This
best-fit model is compared in Table~\ref{table:phi4comp} to the
corresponding model with the full set of single field inflationary
parameters. The $\lambda\phi^4$ model is displaced from the maximum
likelihood generic single field model by $\Delta\chieff=16$
[$\Delta\chieff$(\WMAP) $=14$, $\Delta\chieff$(CBI+ACBAR+2dFGRS) $=2$], where
$\chieff=-2\ln{\cal L}$ and ${\cal L}$ is the likelihood (see
\citet{verde/etal:2003}).  Since the relative likelihood between the
models is $\exp(-8)$,
and the number of degrees of freedom is approximately three,
$\lambda\phi^4$ is disfavored at more than $3\sigma$. The table shows
that adding external data sets does not make a significant difference
to the $\Delta\chieff$ between the models, and the constraint is
primarily coming from \WMAP\ data.

This result holds only for Einstein gravity. When a non-minimal
coupling of the form $\xi\phi^2R$  ($\xi=1/6$ is the conformal
coupling) is added to the Lagrangian, the coupling changes the
dynamics of $\phi$. This model predicts only a tiny amount of
tensor modes \citep{komatsu/futamase:1999,hwang/noh:1998} in agreement
with the data.
\clearpage
\begin{deluxetable}{lccc}
\tablecaption{Goodness-of-fit Comparison for $\lambda\phi^4$ Model
 \label{table:phi4comp}}
\tablewidth{0pt}
\tablehead{ \colhead{Model} & \colhead{$\chieff$ (\WMAP)} &
 \colhead{$\chieff$ (ext+2dFGRS)} & \colhead{Total $\chieff/\nu$ (\cmb+2dFGRS)}}
\startdata
   Best-fit inflation     & 1428  & 36	& 1464/1379 \\ 
   $\lambda\phi^4$ model  & 1442  & 38  & 1480/1382 \\ 
\enddata
\end{deluxetable}
\clearpage


One can perform a similar analysis on any given inflationary model to
see what constraints the data put on it. Rather than attempt this
Herculean task, in the following section we simply use our constraints
on $n_s$, $dn_s/d\ln k$, and $r$ and the predictions of various
classes of single field inflationary models for these parameters in
order to put broad constraints on them.

\subsubsection{Testing a broad class of inflation models} \label{hsr}

Naively, the parameter space in observables spanned by the slow roll
parameters appears to be large. 
We shall show below that ``viable'' slow roll inflation models 
(i.e. 
those that can sustain inflation for a sufficient number of $e$-folds 
to solve cosmological problems) actually occupy significantly 
smaller regions in the parameter space.

\citet{hoffman/turner:2001,kinney:2002a,easther/kinney:2002,hansen/kunz:2002,caprini/hansen/kunz:2003}
have investigated generic predictions of slow roll inflation models by
using a set of inflationary flow equations (see
Appendix~\ref{A:floweq} for a detailed description and definition of
conventions). In particular \citet{kinney:2002a} and
\citet{easther/kinney:2002} use Monte Carlo simulations to extend the
slow roll approximations to fifth order. These authors  find
``attractors'' corresponding to  fixed points (where all derivatives
of the flow parameters vanish); models cluster strongly near the
power-law inflation  predictions, $r=8(1-n_s)$ (see \S~\ref{small}),
and on the zero tensor modes, $r=0$.

Following the method of \citet{kinney:2002a} and
\citet{easther/kinney:2002}, we compute a million realizations of the
inflationary flow equations numerically, truncating the flow equation
hierarchy at eighth order and evaluating the observables to second
order in slow roll using equations
(\ref{eq:r_2})--(\ref{eq:run_2}). We marginalize over the ambiguity of
converting between $\phi$ and $k$, introduced by the details of
reheating and the energy density during inflation by adopting the
Monte Carlo approach of the above authors. The observable quantities
of a given realization of the flow equations are evaluated at a
specific value of $e$-folding, $N$. However, observable quantities are
measured at a specific value of $k$. Therefore, we need to relate $N$
to $k$. This requires detailed modeling of reheating, which carries an
inherent uncertainty. We attempt to marginalize over this by randomly
drawing $N$ values from a uniform distribution $N=[40,70]$.

\begin{figure}
 \plotone{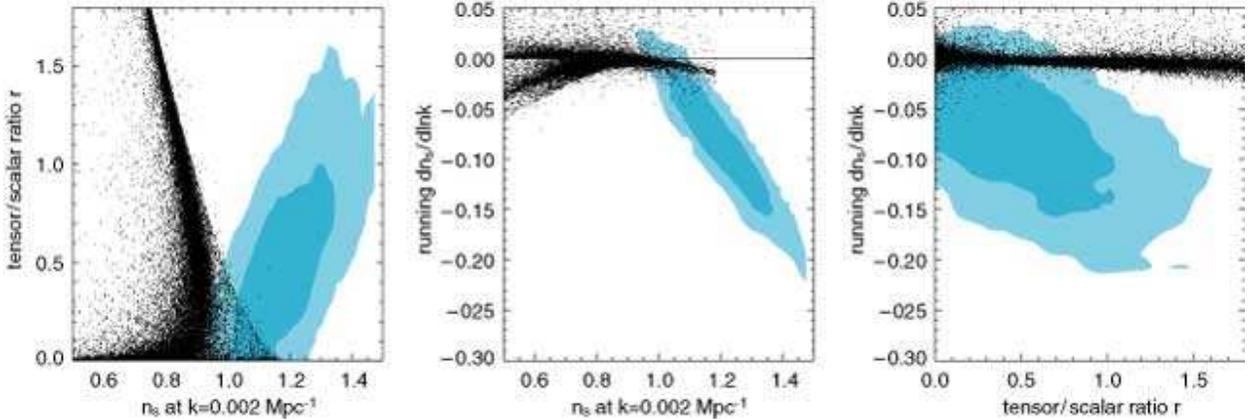}  
\caption{ This set of figures shows part of the
 parameter space spanned by viable slow roll inflation models, with
 the \WMAP\ 68\% confidence region shown in dark blue and the 95\%
 confidence region shown in light blue.}   
\label{fig:obsspace}
\end{figure}

Figure \ref{fig:obsspace} shows part of the parameter space of viable
slow roll inflation models, with the \map 95\% confidence region shown
in blue. Each point on these panels is a different Monte Carlo
realization of the flow equations, and corresponds to a viable slow
roll model. Not all points that are viable slow roll models correspond
to specific physical models constructed in the literature. Most of the
models cluster near the attractors, sparsely populating the rest of
the large parameter space allowed by the slow roll classification. It
must be emphasized that these scatter plots should not be interpreted
in a statistical sense since we do not know how the initial conditions
for the universe are selected. Even if a given realization of the flow
equations does not sit on the attractor, this does \emph{not} mean
that it is not favored. Each point on this plot carries equal weight,
and each is a viable model of inflation. Notice that the \map data do
not lie particularly close to the $r= 8(1-n_s)$ ``attractor''
solution, at the 2-$\sigma$ level, but is quite consistent with the
$r=0$ attractor. 

One may categorize slow roll models into several classes depending
upon where the predictions lie on the parameter space spanned by
$n_s$, $dn_s/d\ln k$, and $r$
\citep{dodelson/kinney/kolb:1997,kinney:1998,hannestad/hansen/villante:2001}.
Each class should correspond to specific physical models of
inflation. Hereafter, we drop the subscript $V$ unless there is an
ambiguity --- it should otherwise be implicitly assumed that we are
referring to the standard slow roll parameters. We categorize the
models on the basis of the curvature of the potential $\eta$, as it is
the only parameter that enters into the relation between $n_s$ and $r$
(equation (\ref{eq:n_s})), and between $n_s$ and $dn_s/d\ln k+2\xi$
(equation (\ref{eq:run_s})).  Thus, $\eta$ is the most important
parameter for classifying the observational predictions of the slow
roll models.  The classes are defined by
\begin{itemize}
  \item[(A)] negative curvature models, $\eta < 0$,
  \item[(B)] small positive (or zero) curvature models, 
             $0\le \eta \le 2\epsilon$, 
  \item[(C)] intermediate positive curvature models, 
             $2\epsilon < \eta \le 3\epsilon$, and
  \item[(D)] large positive curvature models, $\eta > 3\epsilon$.
\end{itemize}
Each class occupies a certain region in the parameter space.
Using $\eta=(n_s-1)\alpha/[2(\alpha-3)]$, where $\eta=\alpha\epsilon$, 
one finds
\begin{itemize}
  \item[(A)] $n_s<1$, $0\le r < \frac83(1-n_s)$, 
             $-\frac23(1-n_s)^2<dn_s/d\ln k+2\xi<0$, 
  \item[(B)] $n_s<1$, $\frac83(1-n_s)\le r \le 8(1-n_s)$,
             $-\frac23(1-n_s)^2\le dn_s/d\ln k+2\xi \le 2(1-n_s)^2$,
  \item[(C)] $n_s<1$, $r>8(1-n_s)$,
             $dn_s/d\ln k+2\xi>2(1-n_s)^2$, and
  \item[(D)] $n_s\ge 1$, $r\ge 0$,
             $dn_s/d\ln k+2\xi>0$.
\end{itemize}
To first order in slow roll, the subspace ($n_s$, $r$) is uniquely
divided into the four classes, and the whole space spanned by these
parameters is defined by this classification.  The division of the
other subspace ($n_s$, $dn_s/\ln k$) is less unique, and $dn_s/d\ln
k<-2\xi - \frac23(1-n_s)^2$ is not covered by this classification. To
higher order in slow roll, these boundaries only hold approximately -
for instance, Case C can have a slightly blue scalar index, and Case D
can have a slightly red one.

We summarize basic predictions of the above model classes to first
order in slow roll using the relation between $r$ and $n_s$ (equation
(\ref{eq:n_s})) rewritten as
\begin{equation}
 \label{eq:r*}
 r=\frac83(1-n_s)+\frac{16}3\eta.
\end{equation}
This implies:
\begin{itemize}
  \item[(A)] negative curvature models predict $\eta < 0$ and $1-n_s>0$; 
             the second term nearly cancels the first to 
             give $r$ too small to detect,
  \item[(B)] small positive curvature models predict $1-n_s>0$ and
             $\eta>0$; a large $r$ is produced,
  \item[(C)] intermediate positive curvature models predict $1-n_s>0$
             and $\eta>0$; a large $r$ is produced, and
  \item[(D)] large positive curvature models predict $1-n_s<0$ and $\eta>0$;
             the first term nearly cancels the second to give $r$ 
             too small to detect.
\end{itemize}
The cancellation of the terms in Case A and Case D implies
$n_s-1\simeq 2\eta$: the steepness of the potential in Case A
and Case D is insignificant compared to the curvature, $\epsilon\ll
\left|\eta\right|$.  On the other hand, in Case B and Case C the
steepness is larger than or comparable to the curvature, by definition;
thus, non-detection of $r$ can exclude many models in Case B and Case C.
As we have shown in \S~\ref{phi4}, a minimally-coupled 
$\lambda\phi^4$ model, which falls into Case B, is excluded 
at high significance,
largely because of our non-detection of $r$ (see also \S~\ref{small}).



\begin{figure}
\plotone{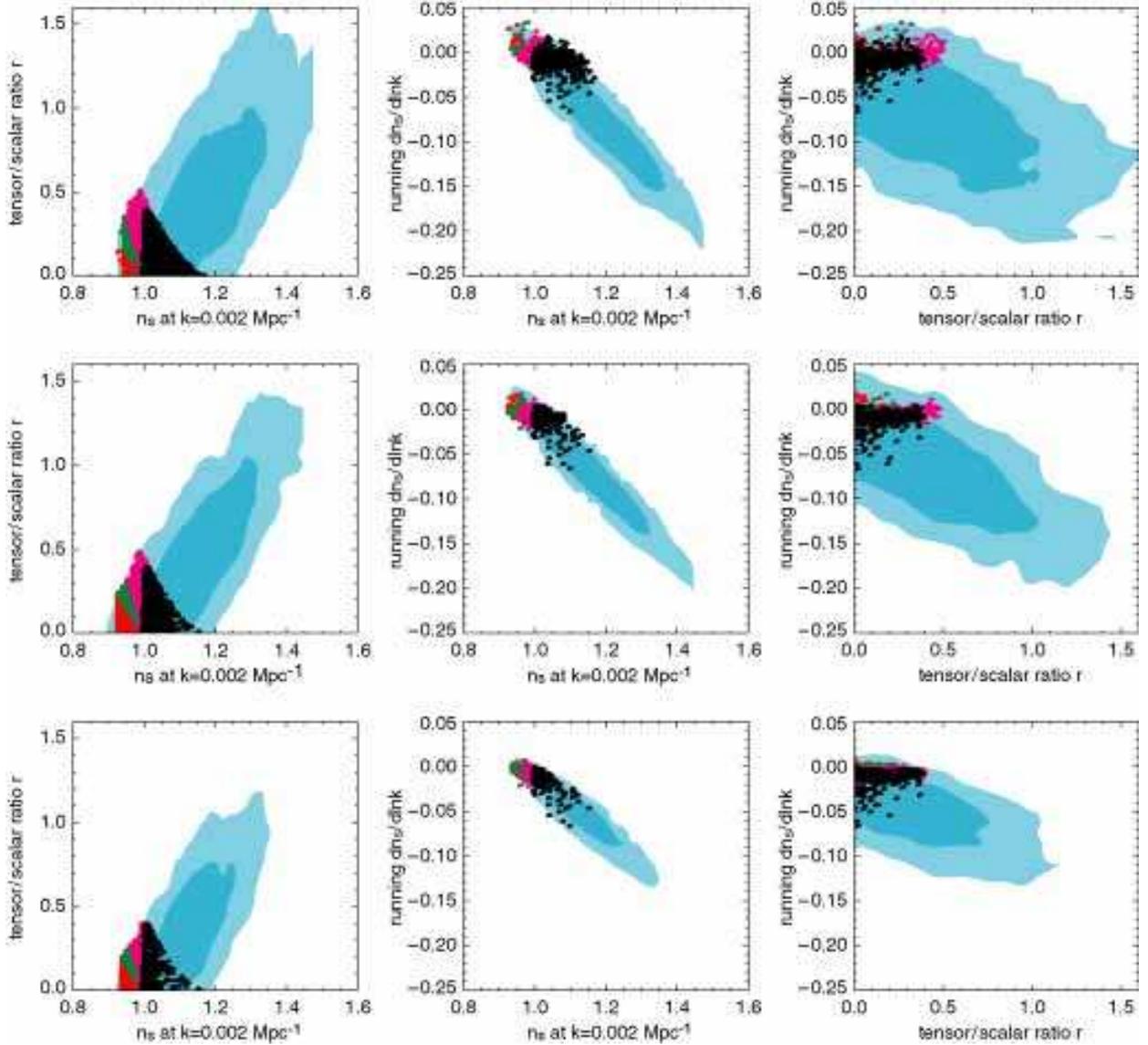} 
\caption{This set of figures compares the fits from
 the \WMAP\ (top row), \cmb+2dFGRS (middle row) and
 \cmb+2dFGRS+Ly$\alpha$ data (bottom row) to the predictions of
 specific classes of physically motivated inflation models. The color
 coding shows model classes referred to in the text: (A) red, (B)
 green, (C) magenta, (D) black. The dark and light blue regions are
 the \emph{joint} 1--$\sigma$ and 2--$\sigma$ regions for the specified data
 sets (contrast this with the 1-d \emph{marginalized} 1--$\sigma$
errors given in Table \ref{table:single_field}).  We show only Monte
Carlo models that are consistent with all three 2--$\sigma$ regions in
each data set. This figure does not imply that the models not plotted
are ruled out.} 
\label{fig:inflcmb2dflyamodel}
\end{figure}

For an overview, Figure \ref{fig:inflcmb2dflyamodel} shows the Monte
 Carlo flow equation realizations corresponding to the model classes
 A--D above on the ($n_s$, $r$), ($n_s$, \run), and ($r$, \run)
 planes, for the \WMAP, \cmb+2dFGRS and \cmb+2dFGRS+Lyman $\alpha$
 data sets.

In Table \ref{table:etaclassconstraints}, we show the ranges taken by
the observables $n_s$, $r$ and \run\ in the Monte Carlo realizations
that remain after throwing out all the points which are outside at
least one of the joint-95\% confidence levels. These points have been
separated into the model classes A--D via their $\eta_\sv$. These
constraints were calculated as follows. First, we find the Monte Carlo
realizations of the flow equations from each model class that fall
inside \emph{all} the joint-95\% confidence levels for a given data
set, separately for the \WMAP, \cmb+2dFGRS and \cmb+2dFGRS+Lyman
$\alpha$ data sets (i.e. the models shown on Figure
\ref{fig:inflcmb2dflyamodel}). Then we find for each model class the
maximum and minimum values predicted for each of the observables
within these subsets. These constraints mean that only those models
(within each class) predicting values for the observables that lie
outside these limits are excluded by these data sets at 95\% CL. Note
that the best-fit model within this parameter space has a $\chieff/\nu
=1464/1379$. Here, recall again that the observables were evaluated to
second order in slow roll in these calculations. This is the reason
that the Class C range in $n_s$ goes slightly blue and the Class D
range in $n_s$ goes slightly red; the divisions of the $\eta_\sv$
classification are only exact to first order in slow roll.

In the following subsections we will discuss in more detail the
constraints on specific physical models that fall into the classes
A--D. For a given class, we will plot \emph{only} the flow equation
realizations falling into that category that are consistent with
the 95\% confidence regions of \emph{all} the planes ($n_s$, $r$), ($n_s$,
 \run) and ($r$, \run). 
\clearpage
\begin{deluxetable}{llll}
\rotate
 \tablecaption{Properties of Inflationary Models Present Within the
 Joint-95\% Confidence Region\tablenotemark{a}
 \label{table:etaclassconstraints}}
\tablewidth{0pt}
\tablehead{
   \colhead{Model} & \colhead{\WMAP} & \colhead{\cmb+2dFGRS} & \colhead{\cmb+2dFGRS$+$Lyman $\alpha$}}
\startdata
   A 
   & $(4\times10^{-6})\tablenotemark{b} \le r \le 0.14$
   & $(2\times10^{-6})\tablenotemark{b} \le r \le 0.19$
   & $(4\times10^{-6})\tablenotemark{b} \le r \le 0.16$  \\    
   & $0.94 \le n_s \le 1.00$
   & $0.93 \le n_s \le 1.00$
   & $0.94 \le n_s \le 1.00$   \\    
   & $-0.02\le\ $\run$\ \le 0.02$
   & $-0.04\le\ $\run$\ \le 0.02$
   & $-0.02\le\ $\run$\ \le 0.004$   \\    
   \hline
   B 
   & $(7\times10^{-3})\tablenotemark{b} \le r \le 0.35$
   & $(7\times10^{-3})\tablenotemark{b} \le r \le 0.32$
   & $(7\times10^{-3})\tablenotemark{b} \le r \le 0.26$\\    
   & $0.94 \le n_s \le 1.01$
   & $0.93 \le n_s \le 1.01$
   & $0.94 \le n_s \le 1.01$  \\    
   & $-0.02\le\ $\run$\ \le 0.02$
   & $-0.04\le\ $\run$\ \le 0.02$
   & $-0.02\le\ $\run$\ \le 0.01$  \\    
   \hline
   C  
   & $(0.003)\tablenotemark{b} \le r \le 0.59$
   & $(0.003)\tablenotemark{b} \le r \le 0.52$
   & $(0.03)\tablenotemark{b} \le r \le 0.46$ \\    
   & $0.95 \le n_s \le 1.02$
   & $0.96 \le n_s \le 1.02$ 
   & $0.97 \le n_s \le 1.02$  \\    
   & $-0.04\le\ $\run$\ \le 0.01$ 
   & $-0.04\le\ $\run$\ \le 0.01$ 
   & $-0.04\le\ $\run$\ \le 0.001$  \\    
   \hline
   D  
   & $0.0 \le r \le 1.10$
   & $0.0 \le r \le 0.89$
   & $(8\times10^{-5})\tablenotemark{b} \le r \le 0.89$\\    
   & $0.99 \le n_s \le 1.28$
   & $1.00 \le n_s \le 1.28$ 
   & $1.00 \le n_s \le 1.28$   \\    
   & $-0.09\le\ $\run$\ \le 0.03$
   & $-0.09\le\ $\run$\ \le 0.01$
   & $-0.09\le\ $\run$\ \le -0.001$  \\    
\enddata
 \tablenotetext{a}{The ranges taken by the predicted observables of slow roll
 models (to second order in slow roll) within the joint 95\% CLs from
 the specified data sets. The model classes are: Case A ($\eta<0$),
 Case B ($0\le \eta \le 2\epsilon$), Case C ($2\epsilon < \eta \le
 3\epsilon$), Case D ($\eta>3\epsilon$).}     
\tablenotetext{b}{The lower value of $r$ does not represent a detection,
but rather the minimal level of tensors predicted by any point
in the Monte Carlo that falls within in this class and is consistent
with the data.  We include the lower limit to help set goals for
future CMB polarization missions.}
\end{deluxetable}
\clearpage

Note that very few models predict a ``bad power law'', or $|$\run$|>0.05$.

\subsubsection{Case A: negative curvature models $\eta < 0$} 
\label{negative}

\begin{figure}
\epsscale{0.9}
\plotone{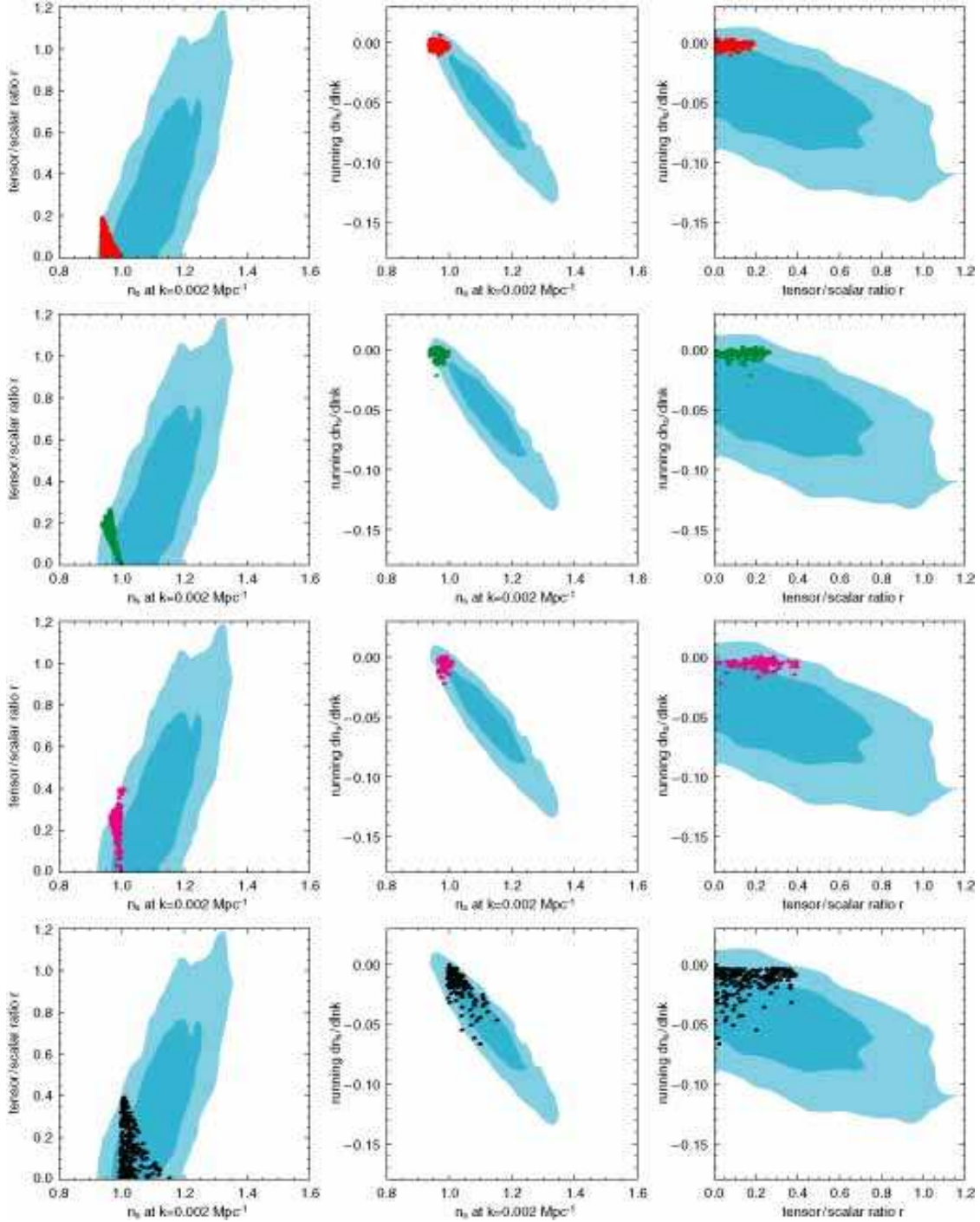}
\caption{This set of figures compares the fits from the
\cmb+2dFGRS+Ly$\alpha$ data to the predictions of all four
classes of inflation models.  The top row is Class A [red dots].
The second row is Class B [green dots].  The third row is
Class C [magenta dots].  The bottom row is Class D [black dots]. 
The dark and light blue regions are the
joint 1--$\sigma$ and 2--$\sigma$ regions for the \cmb+2dFGRS+Ly$\alpha$
data. 
We show only Monte Carlo models that are consistent with
2--$\sigma$ regions in all panels.
This figure does not imply that the models not plotted are ruled out.
\label{fig:f5}}
\end{figure}

The top row of Figure \ref{fig:f5} shows the Monte Carlo points
belonging to Case A which are consistent with all the joint-95\%
confidence regions of the observables shown in the figure, for the
\cmb+2dFGRS+Lyman $\alpha$ data set.  

The negative $\eta$ models often arise from a potential of 
spontaneous symmetry breaking (e.g., new inflation -
\citet{albrecht/steinhardt:1982, linde:1982}).

We consider negative-curvature potentials in the form of
$V=\Lambda^4[1-(\phi/\mu)^p]$ where $p\ge 2$.  We require $\phi<\mu$
for the form of the potential to be valid, and  $\Lambda$ determines
the energy scale of inflation, or the energy stored in a  false
vacuum. One finds that this model always gives a red tilt $n_s<1$ to
first order in slow roll, as $n_s-1=-6\epsilon-2\left|\eta\right|<0$.


For $p=2$, the number of $e$-folds at $\phi$ before the end of
inflation is given by $N\simeq (\mu^2/2\mpl^2)\ln(\mu/\phi)$, where we
have approximated $\phi_{\rm end}\simeq \mu$.  By using the same
approximation, one finds  $n_s-1\simeq -4(\mpl/\mu)^2$, and $r\simeq
32(\phi^2M^2_{\rm pl}/\mu^4)\simeq  8(1-n_s)e^{-N(1-n_s)}$.  In this
class of models, $n_s$ cannot be very close to 1 without $\mu$
becoming larger than $m_{\rm pl}$. For example, $n_s=0.96$ implies
$\mu\simeq 10\mpl\simeq 2m_{\rm pl}$. For this class of models, $r$
has a peak value of $r\simeq 0.06$ at  $n_s=0.98$ (assuming
$N=50$). Even this peak value is too small for {\sl WMAP} to
detect. We see from Table \ref{table:etaclassconstraints} that this
model is consistent with the current data, but requires $\mu>m_{\rm
pl}$ to be valid.

For $p\ge 3$, $n_s-1\simeq -(2/N)(p-1)/(p-2)$ or 
$0.92\le n_s < 0.96$  for $N=50$ regardless of a value of $\mu$, and  $r\simeq
4p^2(\mpl/\mu)^2(\phi/\mu)^{2(p-1)}$ is  negligible as $\phi\ll
\mu$. These models lie in the joint 2--$\sigma$ contour.

The negative $\eta$ model also arises from the potential in the form
of $V=\Lambda^4[1+\alpha\ln(\phi/\mu)]$, a one-loop correction in a
spontaneously broken supersymmetric theory
\citep{dvali/shafi/schaefer:1994}.  Here the coupling constant
$\alpha$ should be smaller than of order 1.  In this model $\phi$
rolls down towards the origin.  One finds $n_s-1=-(1+\frac32\alpha)/N$
which implies $0.95< n_s< 0.98$ for $1> \alpha> 0$ (this formula is
not valid when $\alpha=0$ or $\phi=\mu$).  Since $r=8\alpha/N
=8\alpha(1+\frac32\alpha)^{-1}(1-n_s)= 0.016(\alpha/0.1)$, the tensor
mode is too small for {\sl WMAP} to detect, unless the coupling $\alpha$
takes its maximal value, $\alpha\sim 1$. This type of model is
consistent with the data.

\subsubsection{Case B: small positive curvature models $0\le \eta \le
2\epsilon$}\label{small}

The second row of Figure \ref{fig:f5} shows the Monte Carlo points
belonging to Case B which are consistent with all the joint-95\%
confidence regions of the observables shown in the figure.

The ``small'' positive $\eta$ models correspond to monomial potentials
for $0<\eta<2\epsilon$ and exponential potentials for
$\eta=2\epsilon$.  The monomial potentials take the form of
$V=\Lambda^4 (\phi/\mu)^p$ where $p\ge 2$, and the exponential
potentials $V=\Lambda^4 \exp(\phi/\mu)$.  The zero $\eta$ model is
$V=\Lambda^4 (\phi/\mu)$.  To first order in slow roll, the scalar
spectral index is always red, as $n_s-1=-6\epsilon+2\eta\le
-4\epsilon<0$.  The zero $\eta$ model marks a border between the
negative $\eta$ models and the positive $\eta$ models, giving
$r=\frac83(1-n_s)$.

The monomial potentials often appear in chaotic inflation models
\citep{linde:1983}, which require that $\phi$ be initially displaced
from the origin by a large amount, $\sim m_{\rm pl}$, in order to
avoid fine-tuned initial values for $\phi$. The monomial potentials
can have a period of inflation at $\phi\ga m_{\rm pl}$, and inflation
ends when $\phi$ rolls down to near the origin.  For $p=2$, inflation
is driven by the mass term, which gives  $\phi=2\sqrt{N}\mpl$,
$n_s=1-2/N=0.96$,  $r=8/N=4(1-n_s)=0.16$, and  $dn_s/d\ln
k=-2/N^2=-(1-n_s)^2/2=-0.8\times 10^{-3}$.  For $p=4$, inflation is
driven by the self-coupling, which gives $\phi=2\sqrt{2N}\mpl$,
$n_s=1-3/N=0.94$,  $r=16/N=\frac{16}3(1-n_s)=0.32$, and  $dn_s/d\ln
k=-3/N^2=-(1-n_s)^2/3=-1.2\times 10^{-3}$.  The most striking feature of
the small positive $\eta$ models is  that the gravitational wave
amplitude can be large,  $r\ge 0.16$.   Our data suggest that, for
monomial potentials to lie within the joint 95\% contour, $r < 0.26$
(Table~\ref{table:etaclassconstraints}).  A $\lambda\phi^4$ model is
excluded at $\sim 3$--$\sigma$ (\S~\ref{phi4}), and any monomial
potentials with $p>4$ are also excluded at high signifcance.  Models
with $p = 2$ (mass term inflation) are consistent with the data.

The exponential potentials appear in the Brans--Dicke theory of
gravity \citep{brans/dicke:1961,dicke:1962} conformally transformed 
to the Einstein frame (the extended inflation models) 
\citep{la/steinhardt:1989}. 
One finds $n_s=1-(\mu/\mpl)^2$, $r=8(1-n_s)$, and $dn_s/d\ln k=0$. 
Thus, the exponential potentials predict an exact power-law spectrum
and significant gravitational waves for significantly tilted spectra.
Since $\mu=N \mpl^2 /(\phi-\phi_{end})$,
$n_s=1-[N\mpl/(\phi-\phi_{end})]^2$.
The 95\% range for $n_s$ in Table \ref{table:etaclassconstraints} 
implies that $\phi-\phi_{end} 
> 4N\mpl\simeq 200\mpl \simeq 40 m_{\rm pl}$. 

The exponential potentials mark a border between the small positive 
$\eta$ models and the positive intermediate $\eta$ models described below. 

\subsubsection{Case D: large positive curvature models $\eta > 3\epsilon$}
\label{large}

Before describing Case C, it is useful to describe Case D
first. The fourth row of Figure \ref{fig:f5} shows the Monte Carlo points
belonging to Case D which are consistent with all the joint-95\%
confidence regions of the observables shown in the figure.

The ``large'' positive curvature models correspond to  hybrid
inflation models \citep{linde:1994}, which have recently  attracted
much attention as an $R$-invariant supersymmetric theory
naturally realizes hybrid inflation
\citep{copeland/etal:1994,dvali/shafi/schaefer:1994}.  While it is
pointed out that supergravity effects add too large an effective mass
to the inflaton field to maintain inflation, the minimal K\"ahler
supergravity does not  have such a large mass problem
\citep{copeland/etal:1994,linde/riotto:1997}.  The distinctive feature of
this class of models with $\eta>3\epsilon$ is that the spectrum has a
blue tilt, $n_s-1=-6\epsilon+2\eta>0$, to first order in slow roll.

A typical potential is a monomial potential plus a constant term,
$V=\Lambda^4[1+(\phi/\mu)^p]$, which enables inflation to occur for a
small value of $\phi$, $\phi<m_{\rm pl}$.  At  first sight,
inflation never ends for this potential, as the constant term
sustains the exponential expansion forever.  Hybrid inflation models
postulate a second field $\sigma$ which couples to  $\phi$.  When
$\phi$ rolls slowly on the potential, $\sigma$ stays at the origin and
has no effect on the dynamics.  For a small value of $\phi$ inflation
is dominated by a false vacuum term, $V(\phi,\sigma=0)\simeq
\Lambda^4$.  When $\phi$ rolls down to some critical value, $\sigma$
starts moving toward a true vacuum state, $V(\phi,\sigma)=0$, and
inflation ends.  
A numerical calculation \citep{linde:1994} suggests that the potential
is described by $\phi$ only until $\phi$ reaches a critical value.
When $\phi$ reaches the critical value, inflation suddenly ends
and $\sigma$ need not be considered. Thus, 
we include the hybrid models  in our discussion of 
single-field models.

For $p=2$, one finds that $N\simeq \frac{1}{2}(\mu/M_{\rm
pl})^2\ln(\phi/\phi_{\rm end})\simeq 50$, which, in turn, implies
$\mu\sim 10\mpl\simeq 2m_{\rm pl}$ for $\ln(\phi/\phi_{\rm end})\sim
1$.  The spectral slope is estimated as  $n_s\simeq
1+4(\mpl/\mu)^2\sim 1.04$, and the tensor/scalar ratio, $r\simeq
32(\phi/\mu)^2(\mpl/\mu)^2 =8(\phi/\mu)^2(n_s-1)$, is negligible as
inflation occurs  at $\phi\ll \mu$.  The running is also negligible,
as  $dn_s/d\ln k\simeq 64(\phi/\mu)^2(M_{\rm pl}/\mu)^4=
4(\phi/\mu)^2(n_s-1)^2\ll 10^{-2}$. This type of model lies 
within the joint 95\% contours.

One-loop correction in a softly broken supersymmetric theory induces a
logarithmically running mass,
$V=\Lambda^4\left\{1+(\phi/\mu)^2\left[1+\alpha\ln(\phi/Q)\right]\right\}$,
where $\alpha$ is a coupling constant and $Q$ is a renormalization
point.  Since $n_s$ is practically determined by $V''$, this potential
gives rise to a logarithmic running of $n_s$ \citep{lyth/riotto:1999}.
These models would lie in the region occupied by the Monte Carlo
points that have a large, negative $dn_s/d\ln k$. 
This type of model is consistent with the data. 

\subsubsection{Case C: intermediate positive curvature models 
$2\epsilon < \eta \le 3\epsilon$}  
\label{intermediate}

The third row of Figure \ref{fig:f5} shows the Monte Carlo points
belonging to Case C which are consistent with all the joint-95\%
confidence regions of the observables shown in the figure. 

The ``intermediate'' positive curvature models are defined, to first
order in slow roll, as having a
red tilt, $n_s-1=-6\epsilon+2\eta< 0$, or the exactly scale-invariant
spectrum, $n_s-1=0$, while not being described by monomial or
exponential  potentials. These conditions lead to a parameter space
where $2\epsilon<\eta\le 3\epsilon$. Here we discuss only examples of
physical models that do not solely live in Case C, but briefly pass
through it as they transition from Case D to Case B or Case A.

The transition from Case D to Case B may correspond to a special case
of hybrid inflation models described in the previous subsection (Case
D), $V=\Lambda^4[1+(\phi/\mu)^p]$.  When $\phi\gg \mu$, the potential
becomes Case B potential,  $V\rightarrow \Lambda^4(\phi/\mu)^p$, and
the spectrum is red, $n_s<1$.  When $\phi\ll \mu$, the potential
drives hybrid inflation, and the spectrum is blue, $n_s>1$.
On the other hand, when $\phi\sim \mu$, the potential takes a
parameter space somewhere between Case B and Case D, which corresponds
to Case C.  One may argue that this model requires fine-tuned
properties in that we just transition from one regime to the other.
However, the Case C regime has an interesting property: the spectral index
$n_s$ runs from red on large scales to blue on small scales,  as
$\phi$ undergoes the transition from Case B to Case D. This example
has the wrong sign for the running of the index compared to the data at
the $\sim 2$-$\sigma$ level.

\cite{linde/riotto:1997} is one example of a transition from Case D to
Case A.  They consider a supergravity-motivated hybrid potential with
a  one-loop correction, which can be approximated during inflation as
\begin{equation}
 \label{eq:linderiotto}
  V\simeq \Lambda^4\left[1+\alpha\ln(\phi/Q)+\lambda(\phi/\mu)^4\right].
\end{equation}
Suppose that the one-loop correction is negligible in some early time,
i.e., $\phi\simeq Q$. The spectrum is blue.  (The third term is
practically unimportant,  as inflation is driven by the first term at
this stage.)  If the loop correction becomes important after several
$e$-folds, then $n_s$ changes from blue to red, as the loop correction
gives a red tilt as we saw in \S~\ref{negative}.  
This example is consistent with the data.
The transition (from Case D to Case A) is possible only 
when  $\alpha$ and $Q$
conspire to balance the first term and the second term right at the
scale accessible to our observations. 


\section{MULTIPLE FIELD INFLATION MODELS} \label{multimodels}

\subsection{Framework}

In general, a candidate fundamental theory of particle physics such as
a supersymmetric theory requires not only one, but many other  scalar
fields.  It is thus naturally expected that during inflation there may
exist more than one scalar field that contributes to the dynamics of
inflation.

In most single-field inflation models, the fluctuations
produced have an almost scale-invariant, Gaussian, purely adiabatic
power spectrum whose amplitude is characterized by the comoving
curvature perturbation, $\hat{\mathcal{R}}$, which remains constant on
superhorizon scales.
They also predict tensor perturbations with the consistency condition  in
equation (\ref{eq:n_t}). 

With the addition of multiple fields, the space of possible
predictions widens considerably.  The most distinctive feature is the
generation of entropy, or isocurvature, perturbations between one
field and the other.  The entropy perturbation, $\hat{\mathcal{S}}$,
can violate the conservation of  $\hat{\mathcal{R}}$ on superhorizon
scales,
providing a source for the late-time evolution of $\hat{\mathcal{R}}$
which weakens the single field consistency condition into an upper
bound on the tensor/scalar ratio
\citep{polarski/starobinsky:1995,sasaki/stewart:1996,garcia-bellido/wands:1996}.
Limits on the possible level of the entropy perturbation thus
discriminate between the multiple field models and the single field
models.  In this section, we consider the minimal extension to 
single-field inflation -- a model consisting of two minimally-coupled
scalar fields.

\subsection{Correlated Adiabatic/Isocurvature Fluctuations from
Double-Field Inflation} \label{adiso}

The \map data confirm that pure isocurvature fluctuations do not
dominate the observed CMB anisotropy. 
They predict large-scale temperature anisotropies that are too large
with respect to the measured density fluctuations, and have the 
wrong peak/trough
positions in the temperature and polarization power spectra
 \citep{hu/white:1996,page/etal:2003c}.
The \map observations limit but do not preclude the possibility of correlated
mixtures of isocurvature and adiabatic perturbations, which is a
generic prediction of two-field inflation models.  Both
isocurvature and adiabatic perturbations receive significant
contributions from at least one of the scalar fields to produce the
correlation \citep{langlois:1999, pierpaoli/garcia-bellido/borgani:1999, langlois/riazuelo:2000,
gordon/etal:2001, bartolo/matarrese/riotto:2001,
bartolo/matarrese/riotto:2002, amendola/etal:2002, wands/etal:2002}.
We focus on these mixed models in this section.

Let $\hat{\mathcal{R}}_{rad}$ and $\hat{\mathcal{S}}_{rad}$ be the
curvature and entropy perturbations deep in the radiation era,
respectively. At large scales, the temperature anisotropy is given by 
\citep{langlois:1999}: 
\begin{equation}
 \label{eq:langlois}
  \frac{\Delta T}{T}= 
  \frac{1}{5}\left(\hat{\mathcal{R}}_{rad}-2\hat{\mathcal{S}}_{rad}\right),
\end{equation}
in addition to the integrated Sachs--Wolfe effect.
The entropy perturbation,  $\hat{\mathcal{S}}_{rad}\equiv
\delta\rho_{cdm}/\rho_{cdm}-(3/4)\delta\rho_{\gamma}/\rho_{\gamma}$,
remains constant on large scales until  re-entry into the horizon. If
$\hat{\mathcal{R}}_{rad}$ and $\hat{\mathcal{S}}_{rad}$ have the same
sign (correlated), then the large scale temperature anisotropy is
reduced.  If they have opposite signs (anti-correlated), then the
temperature anisotropy is increased.  \cite{spergel/etal:2003} find
that there is an apparent lack of power at the very largest scales in
the \map data; thus, one of the motivations of this study is to
see whether a correlated  $\hat{\mathcal{S}}_{rad}$ can provide a
better fit to the \map low-$l$ data than a purely adiabatic
model.
 
The evolution of the curvature/entropy perturbations from
horizon-crossing to the radiation-dominated era can be parameterized by
a transfer matrix \citep{amendola/etal:2002},
\begin{equation}
 \label{eq:amendola}
  \left( 
   \begin{array}{c} 
    \hat{\mathcal{R}}_{rad} \\
    \hat{\mathcal{S}}_{rad} 
   \end{array} 
  \right) = \left(   
  \begin{array}{cc} 
   1 & T_{RS} \\ 
   0 & T_{SS} 
  \end{array} 
  \right) \left( 
  \begin{array}{c} 
   \hat{\mathcal{R}}_{\star} \\
   \hat{\mathcal{S}}_{\star} 
  \end{array} 
  \right)_{k=aH}.
\end{equation}
Here, $T_{RR}=1$ and $T_{SR}=0$ because of the physical requirement
that $\hat{\mathcal{R}}$ is conserved for purely adiabatic
perturbations, and that $\hat{\mathcal{R}}$ cannot source
$\hat{\mathcal{S}}$.  All the quantities in
equation (\ref{eq:amendola}) are weakly  scale-dependent, and may be
parameterized by power-laws.
Hence, we write this equation as
\begin{eqnarray}
 \hat{\mathcal{R}}_{rad} &=& A_r k^{n_1} \hat{a_r} + A_s k^{n_3} \hat{a_s}, \\ 
 \hat{\mathcal{S}}_{rad} &=& B k^{n_2} \hat{a_s},
\end{eqnarray}
where $\hat{a_r}$ and $\hat{a_s}$ are independent Gaussian random
variables with unit variance,  $\langle \hat{a_r} \hat{a_s}
\rangle = \delta_{rs}$.  The cross-correlation spectrum is given by
$\Delta^2_{RS}(k)\equiv
(k^3/2\pi^2)\langle\hat{\mathcal{R}}_{rad}\hat{\mathcal{S}}_{rad}\rangle
=A_sBk^{n_2+n_3}$.
One may define the correlation coefficient using an angle $\Delta$  as
\begin{equation}
 \label{eq:cosdelta}
 \cos\Delta \equiv 
 \frac{\langle\hat{\mathcal{R}}_{rad}\hat{\mathcal{S}}_{rad}\rangle}
      {\langle\hat{\mathcal{R}}^2_{rad}\rangle^{1/2}
       \langle\hat{\mathcal{S}}^2_{rad}\rangle^{1/2}}
 = \frac{\mathrm{sign}(B)A_s k^{n_3}}{\sqrt{A^2_r k^{2n_1}+A^2_s k^{2n_3}}},
\end{equation}
where $-1\le \cos\Delta \le 1$.  Thus, in general, six parameters
($A_r$, $A_s$, $\cos\Delta$, $n_1$, $n_2$, $n_3$) are needed to
characterize the double-inflation model with correlated
adiabatic/isocurvature perturbations, while $\cos\Delta$ is
scale-dependent. In order to simplify our analysis, we neglect the
scale-dependence of $\cos\Delta$; thus, $n_1=n_3 \ne n_2$ and
$\cos\Delta=\mathrm{sign}(B)A_s/A$.
The power spectra are written as  $\Delta^2_{\cal R}(k)\equiv (k^3/2\pi^2)\langle\hat{{\cal
R}}_{rad}^2\rangle=(A_r^2+A_s^2)k^{2n_1}  \equiv A^2k^{n_{ad}-1}$, and
$\Delta^2_{\cal S}(k)\equiv (k^3/2\pi^2)\langle\hat{{\cal S}}_{rad}^2\rangle=B^2k^{2n_2} \equiv
A^2f_{iso}^2k^{n_{iso}-1}$.  We have defined $n_{ad}-1\equiv 2n_1$ and
$n_{iso}-1\equiv 2n_2$ to coincide with the standard notation for the
scalar spectral index.  The ``isocurvature fraction'' defined by
$f_{iso}\equiv B/A$ determines the relative amplitude of $\hat{{\cal S}}$ to
$\hat{{\cal R}}$.  The cross-correlation spectrum is then written as
$\Delta^2_{RS}(k)=\cos\Delta \sqrt{\Delta^2_{\cal R}(k)\Delta^2_{\cal S}(k)}
=A^2f_{iso}\cos\Delta\ k^{(n_{ad}+n_{iso})/2-1}$.

The temperature and polarization anisotropies are given by
these power spectra:
\begin{eqnarray}
 \label{eq:Cladi}
  C_l^{ad}&\propto&  A^2\int\frac{dk}{k}\left(\frac{k}{k_0}\right)^{n_{ad}-1} 
  \left[g_{l}^{ad}(k)\right]^2, \\
 \label{eq:Cliso}
  C_l^{iso}&\propto&  A^2 f_{iso}^2 \int\frac{dk}{k}\left(\frac{k}{k_0}\right)^{n_{iso}-1} 
  \left[g_{l}^{iso}(k)\right]^2, \\
 \label{eq:Clcorr}
  C_l^{corr}&\propto& A^2 f_{iso} \cos\Delta\int\frac{dk}{k}\left(\frac{k}{k_0}\right)^{(n_{ad}+n_{iso})/2-1} \left[g_{l}^{ad}(k)g_{l}^{iso}(k)\right],
\end{eqnarray}
and the total anisotropy is
$C_l^{tot}=C_l^{ad}+C_l^{iso}+2C_l^{corr}$.  Here, $g_l(k)$ is the
radiation transfer function appropriate to adiabatic or isocurvature
perturbations of either temperature or polarization anisotropies. Note
that the quantities $n_{ad}$, $n_{iso}$, and $f_{iso}$ are defined at
a specific wavenumber $k_0$, which we take to be $k_0=\mathrm{0.05 \
Mpc^{-1}}$ in the MCMC. To translate the
constraint on $f_{iso}$ to any other wavenumber, one uses
\be
\label{eq:conv_fiso}
f_{iso}(k_1)=f_{iso}(k_0)\left(\frac{k_1}{k_0}\right)^{(n_{iso}-n_{ad})/2}.
\ee
We can restrict $f_{iso}\ge 0$ without loss of generality. Since we
can remove $A$ by normalizing to the overall amplitude of
fluctuations in the \map data, we are left with 4 parameters,
$n_{ad}$, $n_{iso}$, $f_{iso}$, and $\cos\Delta$.  We neglect the
contribution of tensor modes, as the addition of tensors goes in the
opposite direction in terms of explaining the low amplitude of the
low-$l$ TT power spectrum. We also neglect the scale-dependence of
$n_{ad}$ and $n_{iso}$, as they are not well constrained by our data
sets.

\begin{figure}
\plotone{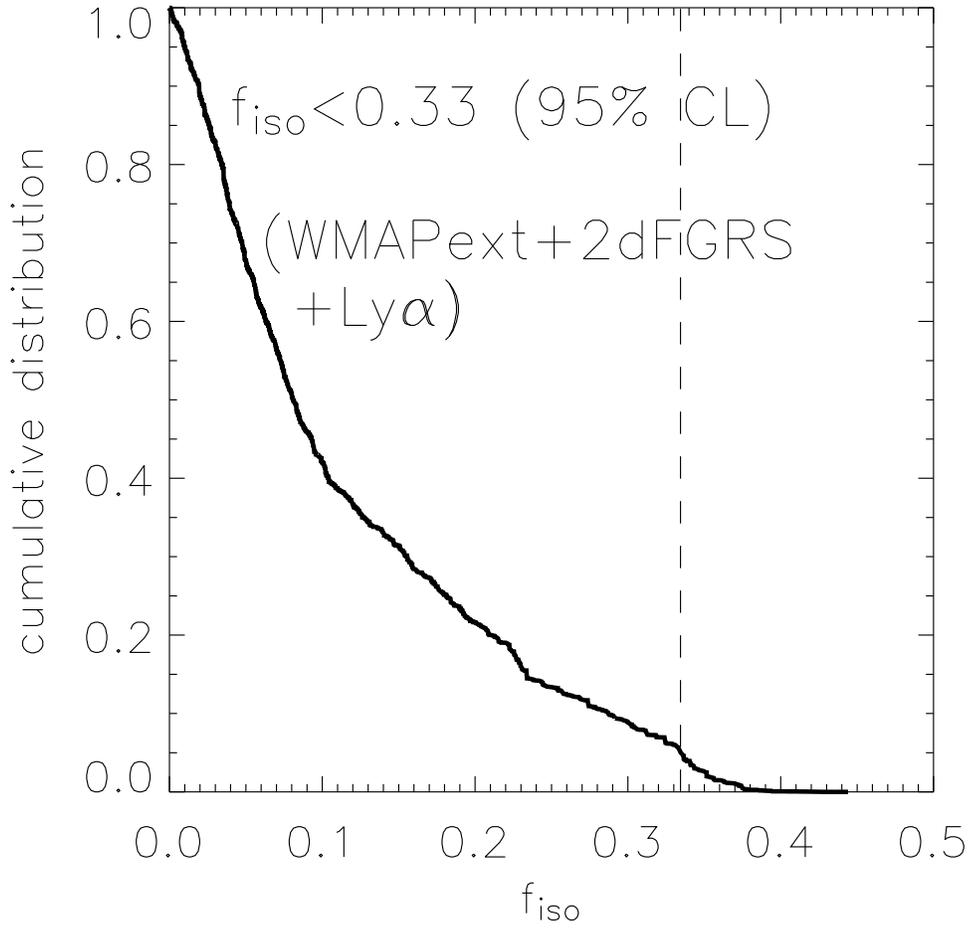} 
\vspace{10pt}
\caption{The cumulative distribution of the isocurvature fraction,
$f_{iso}$, for the \cmb+2dFGRS+Lyman $\alpha$ data set.} 
\label{fig:fiso}
\end{figure}

We fit to the \cmb+2dFGRS and \cmb+2dFGRS+Lyman $\alpha$ data sets with the 11
parameter model ($\Omega_b h^2$, $\Omega_m h^2$, $h$, $\tau$,
$n_{ad}$, $n_{iso}$, $f_{iso}$, $\cos \Delta$, $A$, $\beta$,
$\sigma_p$). The results of the fit for the double inflation model
parameters are shown in
Table~\ref{table:adisoparam}. Figure ~\ref{fig:fiso} shows the cumulative
distribution of $f_{iso}$. The best-fit non-primordial cosmological
parameter constraints are very similar to the single field case. 
\clearpage
\begin{deluxetable}{lcc}
\tablecaption{Cosmological Parameters: Adiabatic + Isocurvature Model
\label{table:adisoparam}}
\tablehead{\colhead{Parameter} &  \colhead{\cmb+2dFGRS}  &
\colhead{\cmb+2dFGRS$+$Lyman $\alpha$}}

\startdata
$f_{iso}(k_0=0.05~{\rm Mpc}^{-1})$ &$<0.32$\tablenotemark{a} & $<0.33$\tablenotemark{a} \\    
$n_{ad}$  & $0.97 \pm 0.03$ & $0.95 \pm 0.03$\\
$n_{iso}$ & $1.26_{-0.57}^{+0.51}$ & $1.29_{-0.56}^{+0.50}$\\
$\cos\Delta$ & $-0.76_{-0.14}^{+0.18}$  & $-0.76_{-0.16}^{+0.18}$\\
$A(k_0=0.05~{\rm Mpc}^{-1})$ & $0.82\pm 0.10$  & $0.78\pm 0.08$ \\
$\Omega_b h^2$ & $0.023 \pm 0.001$  & $0.023 \pm 0.001$  \\
$\Omega_m h^2$ & $0.133 \pm 0.007$  & $0.131 \pm 0.006$ \\
$h$            & $0.072\pm 0.04$    & $0.072\pm0.04$ \\
$\tau$         & $0.16 \pm 0.06$    & $0.14 \pm 0.06$\\
$\sigma_8$     & $0.84 \pm 0.06$    & $0.81 \pm 0.04$
\enddata    
\tablenotetext{a}{The constraint on the isocurvature fraction,
 $f_{iso}$, is a 95\% upper limit.}

\end{deluxetable}
\clearpage

While the fit tries to reduce the large-scale anisotropy with an
admixture of \emph{correlated} isocurvature modes as expected (note
that $\cos\Delta<0$ corresponds to $\hat{\mathcal{R}}_{rad}$ and
$\hat{\mathcal{S}}_{rad}$ having the same sign, from the definition of
initial conditions in the {\sf CMBFAST} code), this only leads to a
small reduction in amplitude at the
quadrupole. Table~\ref{table:chicomp} compares the goodness-of-fit for
this model along with the maximum likelihood models for the $\Lambda$CDM and single field inflation cases.  Because $\chieff/\nu$ is
not improved by the addition of three new parameters
and considerable physical complexity, we conclude that 
the data do not require this model.  
This implies that the initial conditions are
consistent with being fully adiabatic.
\clearpage
\begin{deluxetable} {lc}
\tablecaption{Goodness-of-Fit Comparison for Adiabatic/Isocurvature
Model \label{table:chicomp}}
\tablewidth{0pt}
\tablehead{
\colhead{Model} &\colhead{$\chieff/\nu$\tablenotemark{a}} 
}
\startdata
$\Lambda$CDM           &  1468/1381 \\
Single field inflation &  1464/1379 \\
Adiabatic/Isocurvature &  1468/1378 
\enddata
\tablenotetext{a}{These $\chieff$ values are for the \cmb+2dFGRS data
set. Here we do not give $\chieff$ for the Lyman $\alpha$ data, as
the covariance between the data points is not known
\citep{verde/etal:2003}.}
\end{deluxetable}
\clearpage
\section{SMOOTHNESS OF THE INFLATON POTENTIAL} \label{features}

\citet{spergel/etal:2003} point out that there are several
sharp features in the \map TT angular power spectrum that contribute
to the reduced-$\chieff$ for the best-fit model being $\sim 1.09$. 
The large $\chieff$ may result from neglecting 0.5--1\% contributions
to the \map TT power spectrum covariance matrix; for example,
gravitational lensing of the CMB, beam asymmetry, and non-Gaussianity 
in noise maps.
When included, these effects will
likely improve the reduced-$\chieff$ of the best-fit $\Lambda$CDM model. 
At the moment we cannot attach any astrophysical reality to these features. 
Similar features appear in Monte Carlo simulations.

{\it While we do not claim these glitches are cosmologically
significant},
it is intriguing to consider what they might imply {\it if} they turn out 
to be significant after further scrutiny.

In this section we investigate whether the reduced-$\chieff$ is improved 
by trying to fit one or more of these ``glitches'' with a feature in
the inflationary potential. 
\citet{adams/ross/sarkar:1997} show that a class of
models derived from supergravity theories naturally gives rise to
inflaton potentials with a large number of sudden downward
steps. 
Each step corresponds to a symmetry-breaking phase transition
in a field coupled to the inflaton, since the mass changes suddenly
when each transition occurs. 
If inflation occurred in the manner suggested by these authors,
a spectral feature is expected every 10-15 $e$-folds.
Therefore, one of these features may be visible in the CMB or 
large-scale structure spectra. 

We use the formalism adopted by \citet{adams/cresswell/easther:2001},
and model the step by the potential
\be \label{step}
V_{step}(\phi)=\frac{1}{2}m^2\phi^2\left[1+
c\tanh\left(\frac{\phi-\phi_s}{d}\right)\right], 
\ee
where $\phi$ is the inflaton field, and the potential has a
step starting at $\phi_s$ with amplitude and gradient determined by $c$
and $d$ respectively. In physically realistic models, the presence of
the step does not interrupt inflation, but affects density
perturbations by introducing scale-dependent oscillations.
\citet{adams/cresswell/easther:2001} describe the phenomenology of
these models: the sharper the step, the larger the amplitude and 
longevity of the ``ringing.''  For our calculations of the power
spectrum in these models, we numerically integrate the Klein--Gordon 
equation using the prescription of \citet{adams/cresswell/easther:2001}.   

We also phenomenologically model a dip in the inflaton potential 
using a toy model of a Gaussian dip centered at $\phi_s$ with 
height $c$ and width $d$:
\be
V_{dip}(\phi)=\frac{1}{2}m^2\phi^2\left(1- c\
\exp\left[\frac{(\phi-\phi_s)^2}{2d^2}\right]\right).
\ee
 
We fix the non-primordial cosmological parameters at the maximum
likelihood values for the $\Lambda$CDM model fitted to the \cmb\ data,
[$\Omega_b h^2=0.022$, $\Omega_m h^2=0.13$, $\tau=0.11$, $A=0.74$,
$h=0.72$].  We then run simulated annealing codes for only the three
parameters:  $\phi_s$, $c$, and $d$, for each potential, fitting to
the \map TT and TE data only. For this section, since this model
predicts sharp features in the angular power spectrum, we had to
modify the standard {\sf CMBFAST} splining resolution, splining at $\Delta
l=1$ for $2\le l <50$ and $\Delta l=5$ for $l \ge 50$.

\begin{figure}
\plottwo{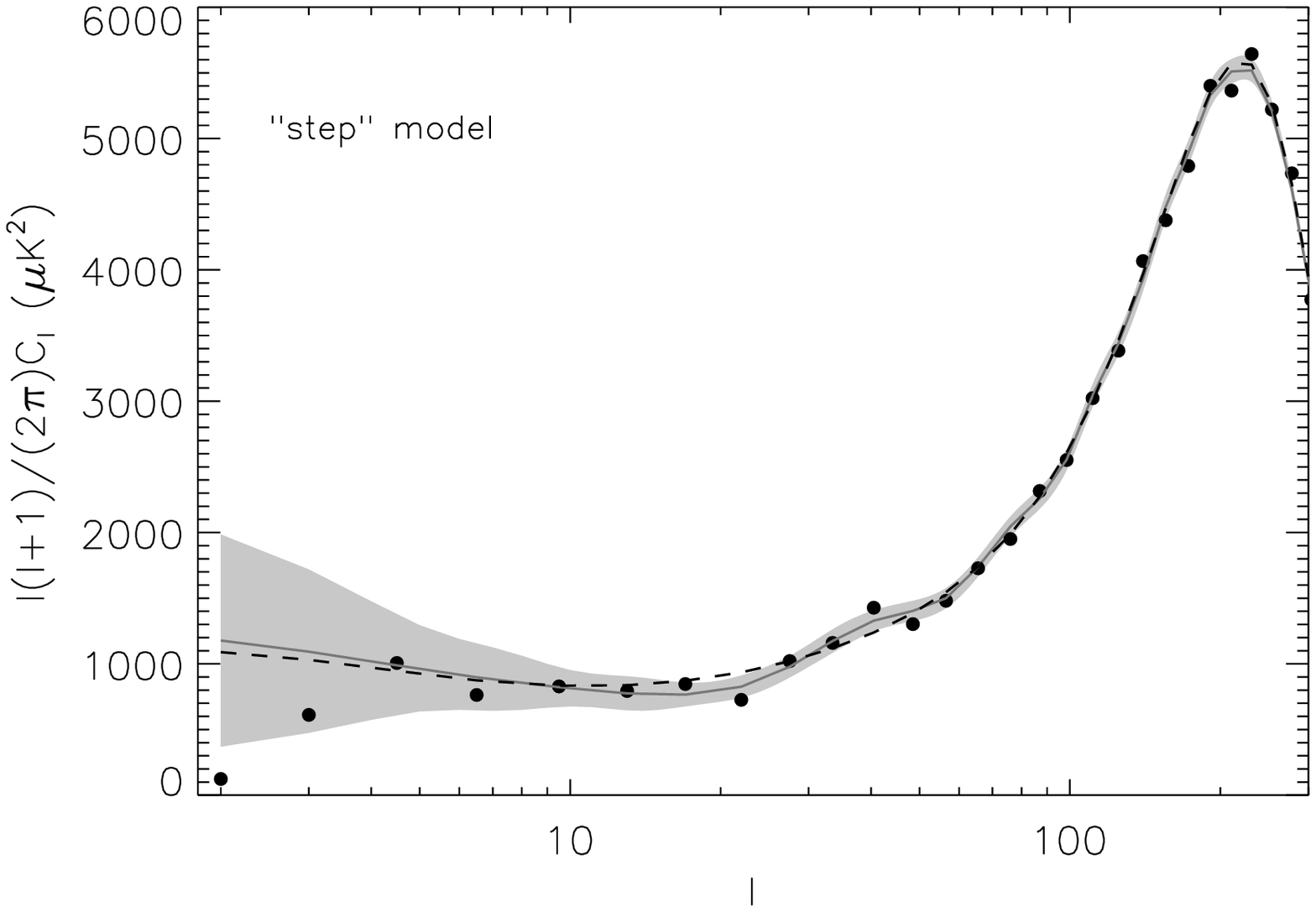}{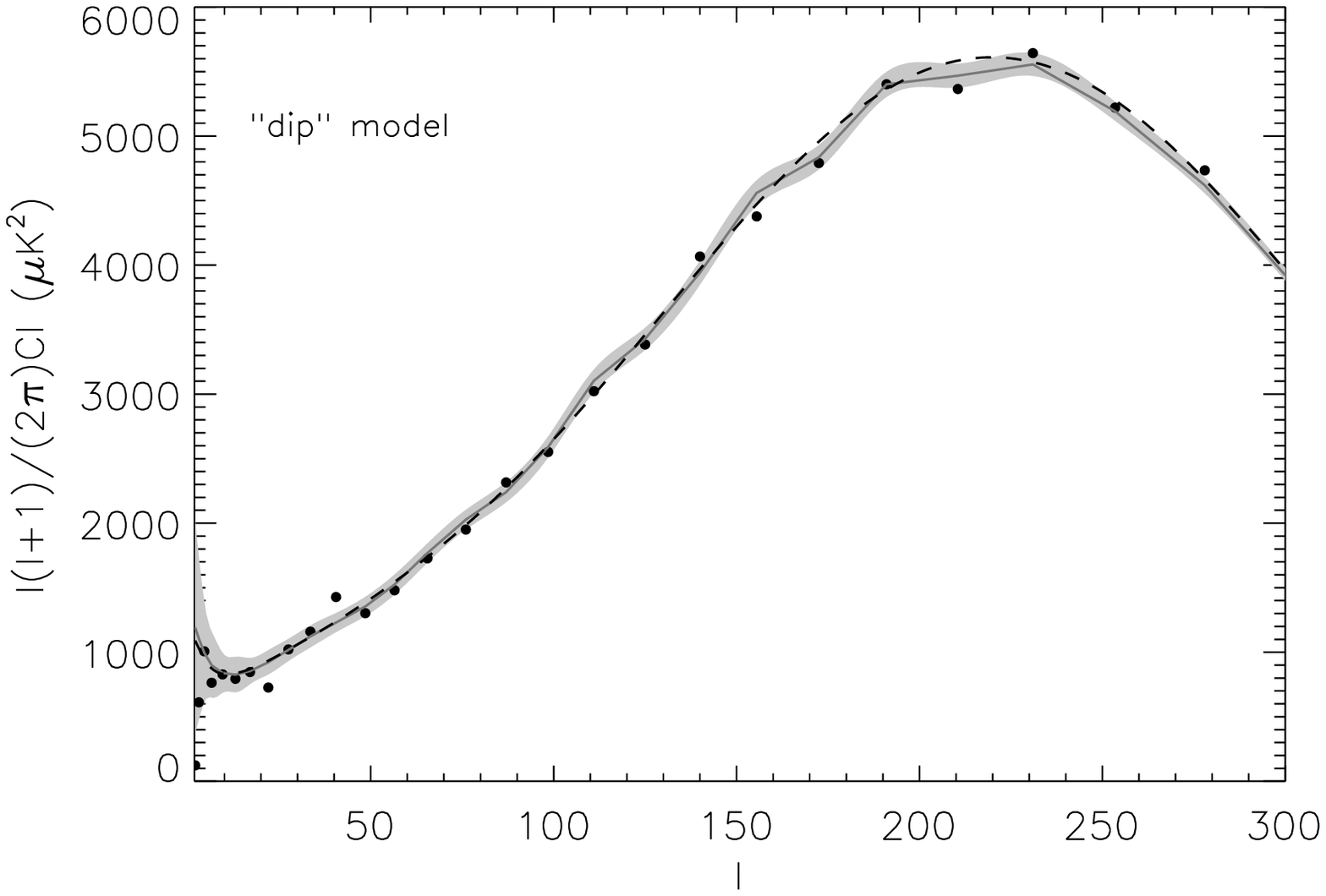} 
\vspace{10pt}
\caption{Best-fit models (solid) with a step (left) and a dip (right) in the
inflaton potential, with the \map TT data. The best-fit $\Lambda$CDM model to \cmb\ data is shown (dotted) for comparison.}
\label{fig:ttstepdip}
\end{figure}

The best-fit parameters found for each potential are given in
Table~\ref{table:features}, along with the $\chieff$ for the \map TT
and TE data. Figure~\ref{fig:ttstepdip} shows these models plotted
along with the \map TT data. 
The best-fit models predict features in the TE spectrum 
at specific multipoles, which are well below detection, given 
the current uncertainties. 
The step model differs from the $\Lambda$CDM model by 
$\Delta \chieff=10$, the dip model by $\Delta\chieff=6$. 
We are not claiming that these are the best possible models in 
this parameter space, only that these are the best-fit models
found in 8 simulated annealing runs.
Note that the models with features were not allowed the freedom to improve
the fit by adjusting the cosmological parameters.  
\clearpage
\begin{deluxetable} {lcccc}
\tablecaption{Best-Fit Models with Potential Features\tablenotemark{a}
\label{table:features}}
\tablehead{
\colhead{Model} &\colhead{$\phi_s$ ($\mpl$)}&\colhead{$c$}&\colhead{$d$ ($\mpl$)}&\colhead{\map $\chieff/\nu$} 
}
\startdata
Step & 15.5379 & 0.00091 &  0.01418 &  1422/1339 \\
Dip  & 15.51757 & 0.00041 &  0.00847 & 1426/1339 \\
$\Lambda$CDM & N/A  & N/A & N/A &  1432/1342 \\
\enddata
\tablenotetext{a}{We give as many significant figures as are needed in
order to reproduce our results.}
\end{deluxetable}
\clearpage

A very small fractional change in the inflaton potential amplitude,
 $c\sim 0.1$\%, is sufficient to cause sharp features in the angular 
power spectrum.
Models with much larger $c$ would have dramatic effects that are not
seen in the \map angular power spectrum.

\begin{figure}
\plottwo{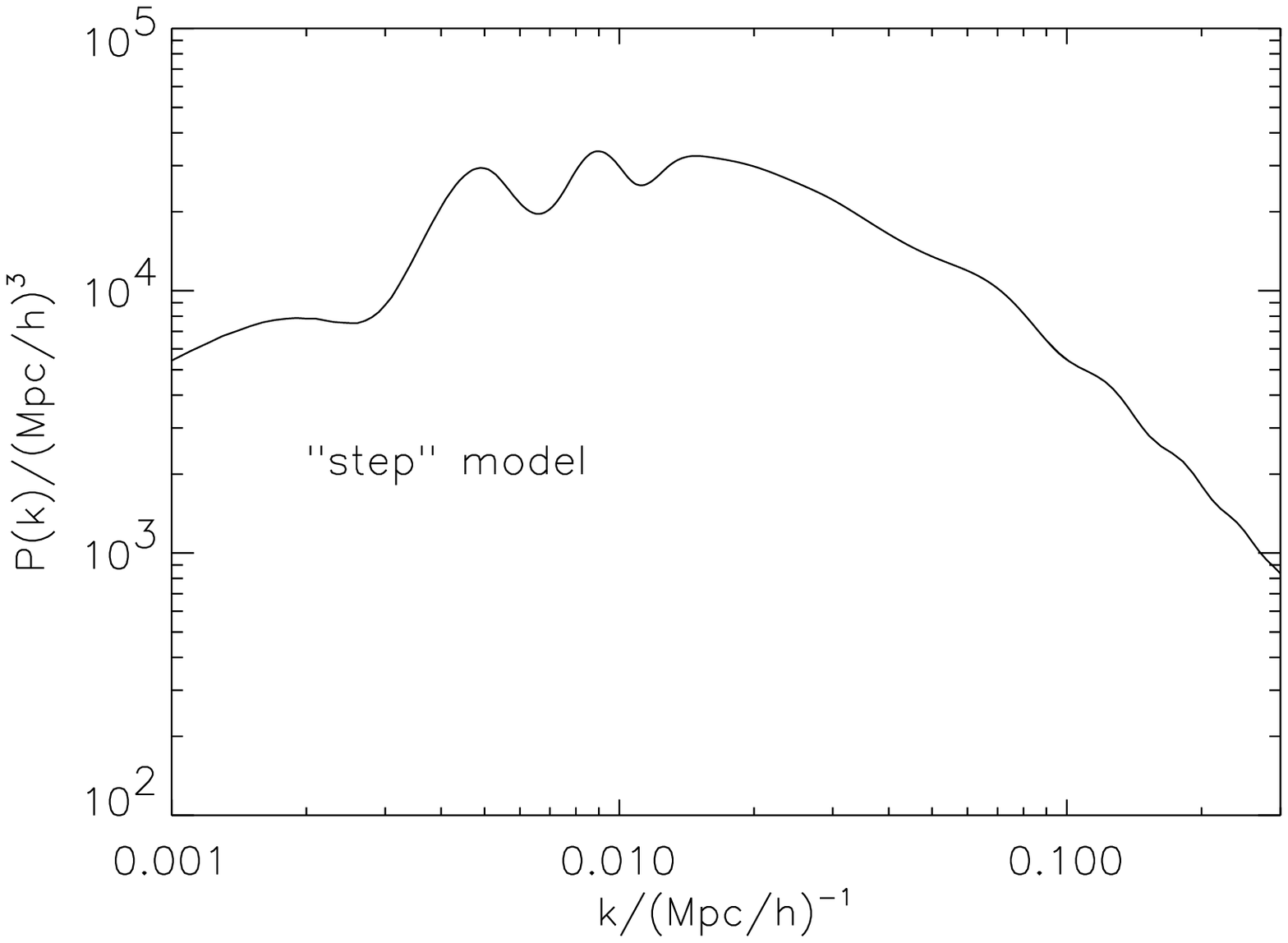}{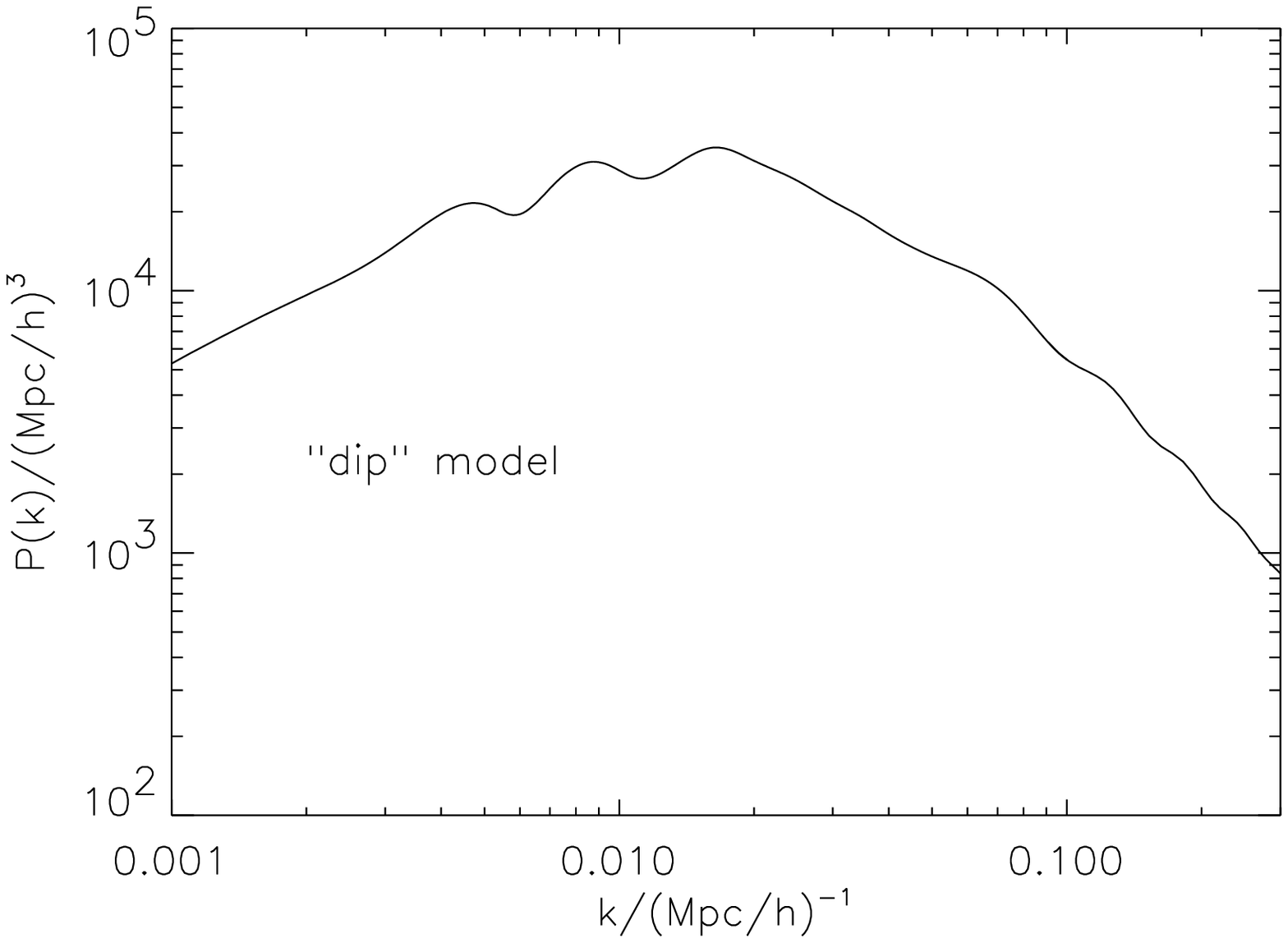} 
\vspace{10pt}
\caption{The large-scale structure power spectra for the best-fit
potential step (left) and dip (right) models.}
\label{fig:stepdipps}
\end{figure}

These models also predict sharp features in the large-scale
structure power spectrum. Figure~\ref{fig:stepdipps} shows the matter
power spectra for the best-fit step/dip models. 
Forthcoming large-scale structure surveys may look for the presence of such
features, and test the viability of these models.   

\section{CONCLUSIONS} \label{finish}
%
%
 \map has made six key observations that are of importance in constraining
 inflationary models.
\begin{itemize}
 \item[(a)] The universe is consistent with being flat 
 \citep{spergel/etal:2003}.
 \item[(b)] The primordial fluctuations are described by random
 Gaussian fields \citep{komatsu/etal:2003}.
 \item[(c)] We have shown that
 the \map detection of an anti-correlation between CMB temperature
 and polarization fluctuations at $\theta > 2^\circ$ is a distinctive
 signature of adiabatic fluctuations on superhorizon
 scales at the epoch of decoupling. This detection agrees with
 a fundamental prediction of the inflationary paradigm.
 \item[(d)] In combination with complementary CMB data (the CBI and
 the ACBAR data), the 2dFGRS large-scale structure data, and Lyman
 $\alpha$ forest data, \map data constrain the primordial scalar and
 tensor power spectra predicted by single-field inflationary
 models. For the scalar modes, the mean and the 68\% error level of the
 1--d marginalized likelihood for the power spectrum slope and 
 the running of the spectral index are, respectively,
 $n_s(k_0=0.002~{\rm Mpc}^{-1})=1.13 \pm 0.08$ and
 \ensuremath{dn_s/d\ln{k} = -0.055^{+ 0.028}_{-0.029}}. This value is
 in agreement with \ensuremath{dn_s/d\ln{k} = -0.031^{+ 0.016}_{-
 0.018}} of \citet{spergel/etal:2003}, which was obtained for a
 $\Lambda$CDM model with no tensors and a running spectral index.
 The data suggest at the 2-$\sigma$ level, but do not require that,
 the scalar spectral index runs from $n_s>1$ on large scales to
 $n_s<1$ on small scales. If true, the third derivative of the
 inflaton potential would be important in describing its dynamics.
 \item[(e)] The \cmb+2dFGRS constraints on $n_s$, \run, and $r$
 put limits on single-field inflationary models that give rise to
 a large tensor contribution and a red ($n_s<1$) tilt.   
 A minimally-coupled $\lambda\phi^4$ model lies more than 3-$\sigma$
 away from the maximum likelihood point. The contribution to the
 $\Delta\chi^2$ between the two points from \map\ alone is 14.
 \item[(f)] We test two-field inflationary models with an admixture of
 adiabatic and CDM isocurvature components. The data do not justify
 adding the additional parameters needed for this model, and the
 initial conditions are consistent with being purely adiabatic.
\end{itemize}


\map both confirms the basic tenets of the inflationary paradigm and
begins to quantitatively test inflationary models. However, we cannot
yet distinguish between broad classes of inflationary theories which
have different physical motivations. In order to go beyond model
building and learn something about the physics of the early universe,
it is important to be able to make such distinctions at high
significance. To accomplish this, one requirement is a better
measurement of the fluctuations at high $l$, and a better measurement
of $\tau$, in order to break the degeneracy between $n_s$ and $\tau$.

We note that an exact scale-invariant spectrum ($n_s=1$ and
$dn_s/d \ln k=0$) is not yet excluded at more than 2$\sigma$ level.
Excluding this point would have profound implications in support of
inflation, as physical single field inflationary models predict
non-zero deviation from exact scale-invariance.

\begin{figure}
\plotone{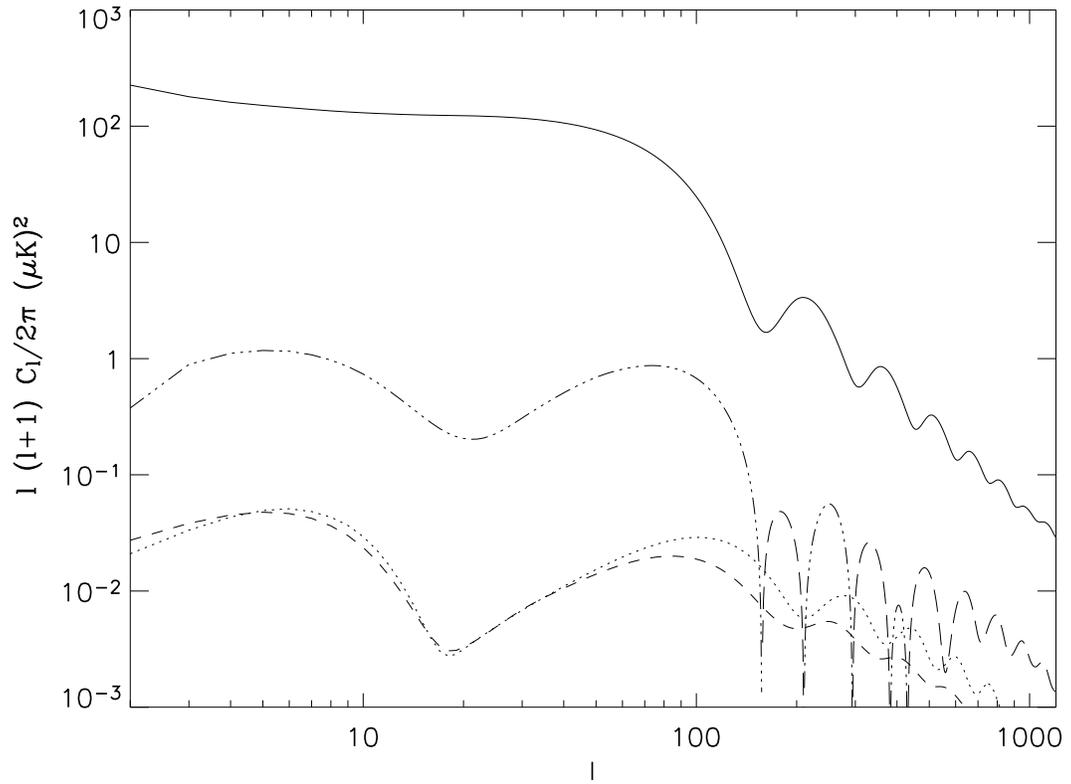} 
\caption{The tensor power spectrum for the maximum likelihood model
from a fit to \cmb+2dFGRS data sets.  The plot shows the TT (solid),
EE (dots), BB (short dashes) and the absolute value of TE negative
(dots and dashes) and positive (long dashes) tensor spectra.}
\label{fig:bmode} 
\end{figure}

We conclude  by showing the tensor temperature and polarization power
spectra for the maximum likelihood single-field inflation model  for
the \cmb+2dFGRS+Lyman $\alpha$ data set, which has tensor/scalar ratio
$r=0.42$ (Figure~\ref{fig:bmode}). The detection and measurement of
the gravity-wave power spectrum would provide the next important key
test of inflation.

\acknowledgements

The \map mission is made possible by the support of the Office of
Space Sciences at NASA Headquarters and by the hard and capable work
of scores of scientists, engineers, technicians, machinists, data
analysts, budget analysts, managers, administrative staff, and
reviewers.   We thank Janet Weiland and Michael Nolta for their
assistance with data analysis and figures.  We thank Uro$\check{\rm
s}$ Seljak for his help with modifications to {\sf CMBFAST}.  HVP
acknowledges the support of a Dodds Fellowship granted by  Princeton
University.  LV is supported by NASA through Chandra Fellowship
PF2-30022 issued by the Chandra X-ray Observatory center, which is
operated by the Smithsonian Astrophysical Observatory for and on
behalf of NASA under contract NAS8-39073.  We thank Martin Kunz for
providing the causal seed simulation results for Figure 1 and Will
Kinney for useful discussions about Monte Carlo simulations of flow
equations.


\appendix 

\section{INFLATIONARY FLOW EQUATIONS} \label{A:floweq}
We begin by describing the hierarchy of inflationary flow equations
described by the generalized ``Hubble Slow Roll'' (HSR) parameters. In
the Hamilton-Jacobi formulation of inflationary dynamics, one
expresses the Hubble parameter directly as a function of the field
$\phi$ rather than a function of time, $H \equiv H(\phi)$, under the
assumption that $\phi$ is monotonic in time. Then the equations of
motion for the field and background are given by:
\begin{eqnarray}
\dot{\phi} &=&-2 \mpl^2 H'(\phi), \label{eq:phih}\\
\left[H'(\phi)\right]^2 -
\frac{3}{2\mpl^2}H^2(\phi)&=&-\frac{1}{2\mpl^4}V(\phi). \label{eq:hj}
\end{eqnarray}
Here, prime denotes derivatives with respect to $\phi$. Equation
(\ref{eq:hj}), referred to as the {\sl Hamilton-Jacobi Equation},
allows us to consider inflation in terms of $H(\phi)$ rather than
$V(\phi)$. The former, being a geometric quantity, describes inflation
more naturally. Given $H(\phi)$, equation (\ref{eq:hj}) immediately
gives $V(\phi)$, and one obtains $H(t)$ by using equation
(\ref{eq:phih}) to convert between $H'$ and $\dot{H}$. This can then
be integrated to give $a(t)$ if desired, since
$H(t)\equiv\dot{a}/a$. Rewriting equation (\ref{eq:hj}) as
\be
H^2(\phi)\left[1-\frac13\epsilon_\sh\right]=\frac{1}{3\mpl^2}V(\phi),
\ee
we obtain
\begin{eqnarray}
\left(\frac{\ddot{a}}{a}\right)&=&\frac{1}{3\mpl^2}[V(\phi)-\dot{\phi}^2] \nonumber\\
                    &=&H^2(\phi)[1-\epsilon_\sh(\phi)], \nonumber
\end{eqnarray}
so that the condition for inflation $(\ddot{a}/a)>0$ is simply given by
$\epsilon_\sh<1$. 

Thus, one can define a set of HSR parameters in analogy to the PSR
parameters of \S~\ref{psr}, though there is no assumption of slow-roll
implicit in this definition:
\begin{eqnarray}
 \label{eq:eps-hsr}
  \epsilon_\sh &\equiv& 2 \mpl^2 \left(\frac{H'(\phi)}{H(\phi)}\right)^2\\ 
 \label{eq:eta-hsr}
  \eta_\sh     &\equiv& 2 \mpl^2 \left(\frac{H''(\phi)}{H(\phi)}\right)\\
 \label{eq:xi-hsr}
  \xi_\sh    &\equiv& 4 \mpl^4 \left(\frac{H'(\phi)H'''(\phi)}{H^2(\phi)}\right)\\
  ^{\ell}\lambda_\sh &\equiv& \left(2 \mpl\right)^{\ell}
  \frac{(H')^{\ell-1}}{H^{\ell}}
  \frac{d^{(\ell+1)}H}{d\phi^{(\ell+1)}}. \label{eq:lam-hsr}
\end{eqnarray}
We need one more ingredient; the number of $e$-folds before the end of
inflation, $N$ is defined by,
\be
\label{eq:Ndef}
N\equiv\int_t^{t_e}H\ dt = \int_\phi^{\phi_e}\frac{H}{\dot{\phi}}
\ d\phi=\frac{1}{\sqrt{2}\mpl}\int^\phi_{\phi_e}\frac{d\phi}{\sqrt{\epsilon_\sh(\phi)}},
\ee
where $t_e$ and $\phi_e$ are the time and field value at the end of
inflation, and $N$ increases the earlier one goes back in time ($t>0
\Rightarrow dN<0$). The derivative with respect to $N$ is therefore, 
\be
\frac{d}{dN}=\frac{\mpl}{2}\sqrt{\epsilon}\frac{d}{d\phi}.
\ee
Then, an infinite hierarchy of inflationary ``flow'' equations can be
defined by differentiating equations (\ref{eq:eps-hsr})--(\ref{eq:lam-hsr})
with respect to $N$:
\begin{eqnarray} 
\frac{d\epsilon_\sh}{dN}&=&2\epsilon_\sh(\eta_\sh-\epsilon_\sh)\\
\frac{d(^{\ell}\lambda_\sh)}{dN}&=&\begin{array}{cr} \left[(\ell-1)\eta_\sh-\ell
\epsilon_\sh\right](^{\ell}\lambda_\sh)+\ ^{\ell+1}\lambda_\sh & (\ell>0) \end{array}.
\end{eqnarray}
The definition of the scalar and tensor power spectra are:
\begin{eqnarray}
\Delta^2_{\cal R} &=& \left[\left(\frac{H}{\dot{\phi}}\right)\left(\frac{H}{2\pi}\right)\right]^2_{k=aH}\\
\Delta^2_h &=& \frac{8}{\mpl^2}\left(\frac{H}{2\pi}\right)^2_{k=aH}.
\end{eqnarray}
Since derivatives with respect to wavenumber $k$ can be expressed with
respect to $N$ as:
\be
\frac{d}{dN}=-(1-\epsilon_\sh)\frac{d}{d \ln k},
\ee
the observables are given in terms of the HSR parameters to second
order as \citep{stewart/lyth:1993,liddle/parsons/barrow:1994}, 
\begin{eqnarray}
r &=& 16\epsilon_\sh\left[1+2C(\epsilon_\sh-\eta_\sh)\right] \label{eq:r_2}\\
n_s-1 &=& (2\eta_\sh-4\epsilon_\sh)\left[1-\frac14(3-5C)\epsilon_\sh\right]-(3-5C)\epsilon_\sh^2+\frac{1}{2}(3-C)\xi_\sh \label{eq:n_2}\\
\frac{dn_s}{d\ln k}
&=&-\left(\frac{1}{1-\epsilon_\sh}\right)\frac{dn_s}{dN} \label{eq:run_2},
\end{eqnarray}
where $C \equiv 4(\ln2+\gamma)-5$ and $\gamma \simeq 0.577$ is Euler's
constant. Note that, as pointed out in \citet{kinney:2002b}, there is
a typographical error in defining $C$ in
\citet{liddle/parsons/barrow:1994} that was inherited by
\citet{kinney:2002a}. We have used the correct value from
\citet{stewart/lyth:1993}.  

Finally, the PSR parameters are given in terms of the HSR parameters
to first order in slow roll as: 
\begin{eqnarray}
\epsilon_\sh&=&\epsilon_\sv\\
\eta_\sh&=&\eta_\sv-\epsilon_\sv\\
\xi_\sh&=&\xi_\sv-3\epsilon_\sv\eta_\sv+3\epsilon_\sv^2.
\end{eqnarray}

\clearpage


\begin{thebibliography}{86}
\expandafter\ifx\csname natexlab\endcsname\relax\def\natexlab#1{#1}\fi

\bibitem[{Adams et~al.(2001)Adams, Cresswell, \&
  Easther}]{adams/cresswell/easther:2001}
Adams, J., Cresswell, B., \& Easther, R. 2001, Phys. Rev., D64, 123514

\bibitem[{Adams et~al.(1997)Adams, Ross, \& Sarkar}]{adams/ross/sarkar:1997}
Adams, J.~A., Ross, G.~G., \& Sarkar, S. 1997, Phys. Lett., B391, 271

\bibitem[{Albrecht et~al.(1996)Albrecht, Coulson, Ferreira, \&
  Magueijo}]{albrecht/etal:1996}
Albrecht, A., Coulson, D., Ferreira, P., \& Magueijo, J. 1996, Phys. Rev.
  Lett., 76, 1413

\bibitem[{{Albrecht} \& {Steinhardt}(1982)}]{albrecht/steinhardt:1982}
{Albrecht}, A. \& {Steinhardt}, P.~J. 1982, \prl, 48, 1220

\bibitem[{Amendola et~al.(2002)Amendola, Gordon, Wands, \&
  Sasaki}]{amendola/etal:2002}
Amendola, L., Gordon, C., Wands, D., \& Sasaki, M. 2002, Phys. Rev. Lett., 88,
  211302

\bibitem[{Bardeen et~al.(1983)Bardeen, Steinhardt, \&
  Turner}]{bardeen/steinhardt/turner:1983}
Bardeen, J.~M., Steinhardt, P.~J., \& Turner, M.~S. 1983, \prd, 28, 679

\bibitem[{Bartolo et~al.(2001)Bartolo, Matarrese, \&
  Riotto}]{bartolo/matarrese/riotto:2001}
Bartolo, N., Matarrese, S., \& Riotto, A. 2001, Phys. Rev., D64, 123504

\bibitem[{Bartolo et~al.(2002)Bartolo, Matarrese, \&
  Riotto}]{bartolo/matarrese/riotto:2002}
---. 2002, Phys. Rev., D65, 103505

\bibitem[{Birrell \& Davies(1982)}]{birrell/davies:1982}
Birrell, N.~D. \& Davies, P. C.~W. 1982, Quantum fields in curved space
  (Cambridge University Press)

\bibitem[{{Brans} \& {Dicke}(1961)}]{brans/dicke:1961}
{Brans}, C. \& {Dicke}, R.~H. 1961, Physical Review, 124, 925

\bibitem[{{Caprini} et~al.(2003){Caprini}, {Hansen}, \&
  {Kunz}}]{caprini/hansen/kunz:2003}
{Caprini}, C., {Hansen}, S.~H., \& {Kunz}, M. 2003, \mnras, 339, 212

\bibitem[{Copeland et~al.(1994)Copeland, Liddle, Lyth, Stewart, \&
  Wands}]{copeland/etal:1994}
Copeland, E.~J., Liddle, A.~R., Lyth, D.~H., Stewart, E.~D., \& Wands, D. 1994,
  Phys. Rev., D49, 6410

\bibitem[{{Croft} et~al.(2002){Croft}, {Weinberg}, {Bolte}, {Burles},
  {Hernquist}, {Katz}, {Kirkman}, \& {Tytler}}]{croft/etal:2002}
{Croft}, R.~A.~C., {Weinberg}, D.~H., {Bolte}, M., {Burles}, S., {Hernquist},
  L., {Katz}, N., {Kirkman}, D., \& {Tytler}, D. 2002, \apj, 581, 20

\bibitem[{{Dicke}(1962)}]{dicke:1962}
{Dicke}, R.~H. 1962, Physical Review, vol.~125, Issue 6, pp.~2163-2167, 125,
  2163

\bibitem[{Dodelson et~al.(1997)Dodelson, Kinney, \&
  Kolb}]{dodelson/kinney/kolb:1997}
Dodelson, S., Kinney, W.~H., \& Kolb, E.~W. 1997, Phys. Rev., D56, 3207

\bibitem[{Durrer et~al.(2002)Durrer, Kunz, \&
  Melchiorri}]{durrer/kunz/melchiorri:2002}
Durrer, R., Kunz, M., \& Melchiorri, A. 2002, Phys. Rept., 364, 1

\bibitem[{Dvali et~al.(1994)Dvali, Shafi, \&
  Schaefer}]{dvali/shafi/schaefer:1994}
Dvali, G.~R., Shafi, Q., \& Schaefer, R. 1994, Phys. Rev. Lett., 73, 1886

\bibitem[{Easther \& Kinney(2002)}]{easther/kinney:2002}
Easther, R. \& Kinney, W.~H. 2002, \prd, submitted (astro-ph/0210345)

\bibitem[{Garcia-Bellido \& Wands(1996)}]{garcia-bellido/wands:1996}
Garcia-Bellido, J. \& Wands, D. 1996, Phys. Rev., D53, 5437

\bibitem[{Gasperini \& Veneziano(1993)}]{gasperini/veneziano:1993}
Gasperini, M. \& Veneziano, G. 1993, Astropart. Phys., 1, 317

\bibitem[{{Gnedin} \& {Hamilton}(2002)}]{gnedin/hamilton:2002}
{Gnedin}, N.~Y. \& {Hamilton}, A.~J.~S. 2002, \mnras, 334, 107

\bibitem[{Gordon et~al.(2001)Gordon, Wands, Bassett, \&
  Maartens}]{gordon/etal:2001}
Gordon, C., Wands, D., Bassett, B.~A., \& Maartens, R. 2001, Phys. Rev., D63,
  023506

\bibitem[{{Gratton} et~al.(2003){Gratton}, {Khoury}, {Steinhardt}, \&
  {Turok}}]{gratton/etal:2003}
{Gratton}, S., {Khoury}, J., {Steinhardt}, P., \& {Turok}, N. 2003, preprint
  (astro-ph/0301395)

\bibitem[{Guth(1981)}]{guth:1981}
Guth, A.~H. 1981, \prd, 23, 347

\bibitem[{Guth \& Pi(1982)}]{guth/pi:1982}
Guth, A.~H. \& Pi, S.~Y. 1982, Phys. Rev. Lett., 49, 1110

\bibitem[{{Hannestad} et~al.(2001){Hannestad}, {Hansen}, \&
  {Villante}}]{hannestad/hansen/villante:2001}
{Hannestad}, S., {Hansen}, S.~H., \& {Villante}, F.~L. 2001, Astroparticle
  Physics, 16, 137

\bibitem[{{Hansen} \& {Kunz}(2002)}]{hansen/kunz:2002}
{Hansen}, S.~H. \& {Kunz}, M. 2002, \mnras, 336, 1007

\bibitem[{Hawking(1982)}]{hawking:1982}
Hawking, S.~W. 1982, Phys. Lett., B115, 295

\bibitem[{{Hinshaw} et~al.(2003)}]{hinshaw/etal:2003}
{Hinshaw}, G.~F. et~al. 2003, \apj, submitted

\bibitem[{Hoffman \& Turner(2001)}]{hoffman/turner:2001}
Hoffman, M.~B. \& Turner, M.~S. 2001, Phys. Rev., D64, 023506

\bibitem[{{Hu} \& {Sugiyama}(1995)}]{hu/sugiyama:1995}
{Hu}, W. \& {Sugiyama}, N. 1995, \apj, 444, 489

\bibitem[{{Hu} \& {White}(1996)}]{hu/white:1996}
{Hu}, W. \& {White}, M. 1996, \apj, 471, 30

\bibitem[{{Hu} \& {White}(1997)}]{hu/white:1997}
---. 1997, \prd, 56, 596

\bibitem[{{Hwang} \& {Noh}(1998)}]{hwang/noh:1998}
{Hwang}, J. \& {Noh}, H. 1998, Physical Review Letters, Volume 81, Issue 24,
  December 14, 1998, pp.5274-5277, 81, 5274

\bibitem[{Khoury et~al.(2002)Khoury, Ovrut, Seiberg, Steinhardt, \&
  Turok}]{khoury/etal:2002}
Khoury, J., Ovrut, B.~A., Seiberg, N., Steinhardt, P.~J., \& Turok, N. 2002,
  Phys. Rev., D65, 086007

\bibitem[{Khoury et~al.(2001)Khoury, Ovrut, Steinhardt, \&
  Turok}]{khoury/etal:2001}
Khoury, J., Ovrut, B.~A., Steinhardt, P.~J., \& Turok, N. 2001, Phys. Rev.,
  D64, 123522

\bibitem[{Kinney(1998)}]{kinney:1998}
Kinney, W.~H. 1998, Phys. Rev., D58, 123506

\bibitem[{Kinney(2002{\natexlab{a}})}]{kinney:2002a}
---. 2002{\natexlab{a}}, Phys. Rev., D66, 083508

\bibitem[{Kinney(2002{\natexlab{b}})}]{kinney:2002b}
---. 2002{\natexlab{b}}, preprint (astro-ph/0206032)

\bibitem[{Kogut et~al.(2003)}]{kogut/etal:2003}
Kogut, A. et~al. 2003, \apj, submitted

\bibitem[{Komatsu \& Futamase(1999)}]{komatsu/futamase:1999}
Komatsu, E. \& Futamase, T. 1999, Phys. Rev., D59, 064029

\bibitem[{Komatsu et~al.(2003)}]{komatsu/etal:2003}
Komatsu, E. et~al. 2003, \apj, submitted

\bibitem[{Kuo et~al.(2002)}]{kuo/etal:2002}
Kuo, C.~L. et~al. 2002, \apj, astro-ph/0212289

\bibitem[{La \& Steinhardt(1989)}]{la/steinhardt:1989}
La, D. \& Steinhardt, P.~J. 1989, Phys. Rev. Lett., 62, 376

\bibitem[{Langlois(1999)}]{langlois:1999}
Langlois, D. 1999, Phys. Rev., D59, 123512

\bibitem[{Langlois \& Riazuelo(2000)}]{langlois/riazuelo:2000}
Langlois, D. \& Riazuelo, A. 2000, Phys. Rev., D62, 043504

\bibitem[{{Leach} et~al.(2002){Leach}, {Liddle}, {Martin}, \&
  {Schwarz}}]{leach/etal:2002}
{Leach}, S.~M., {Liddle}, A.~R., {Martin}, J., \& {Schwarz}, D.~J. 2002, \prd,
  66, 23515

\bibitem[{{Lewis} et~al.(2000){Lewis}, {Challinor}, \&
  {Lasenby}}]{lewis/challinor/lasenby:2000}
{Lewis}, A., {Challinor}, A., \& {Lasenby}, A. 2000, \apj, 538, 473

\bibitem[{Liddle \& Lyth(1992)}]{liddle/lyth:1992}
Liddle, A.~R. \& Lyth, D.~H. 1992, Phys. Lett., B291, 391

\bibitem[{Liddle \& Lyth(1993)}]{liddle/lyth:1993}
---. 1993, Phys. Rept., 231, 1

\bibitem[{Liddle \& Lyth(2000)}]{liddle/lyth:CIALSS}
---. 2000, Cosmological inflation and large-scale structure (Cambridge
  University Press)

\bibitem[{Liddle et~al.(1994)Liddle, Parsons, \&
  Barrow}]{liddle/parsons/barrow:1994}
Liddle, A.~R., Parsons, P., \& Barrow, J.~D. 1994, Phys. Rev., D50, 7222

\bibitem[{Linde(1982)}]{linde:1982}
Linde, A.~D. 1982, Phys. Lett., B108, 389

\bibitem[{Linde(1983)}]{linde:1983}
---. 1983, Phys. Lett., B129, 177

\bibitem[{Linde(1990)}]{linde:1990}
---. 1990, Particle physics and inflationary cosmology (Chur, Switzerland:
  Harwood)

\bibitem[{Linde(1994)}]{linde:1994}
---. 1994, Phys. Rev., D49, 748

\bibitem[{Linde \& Riotto(1997)}]{linde/riotto:1997}
Linde, A.~D. \& Riotto, A. 1997, Phys. Rev., D56, 1841

\bibitem[{Lyth \& Riotto(1999)}]{lyth/riotto:1999}
Lyth, D.~H. \& Riotto, A. 1999, Phys. Rept., 314, 1

\bibitem[{{Magueijo} et~al.(1996){Magueijo}, {Albrecht}, {Coulson}, \&
  {Ferreira}}]{magueijo/etal:1996}
{Magueijo}, J., {Albrecht}, A., {Coulson}, D., \& {Ferreira}, P. 1996, \prl,
  76, 2617

\bibitem[{{Mukhanov} \& {Chibisov}(1981)}]{mukhanov/chibisov:1981}
{Mukhanov}, V.~F. \& {Chibisov}, G.~V. 1981, JETP Letters, 33, 532

\bibitem[{Mukhanov et~al.(1992)Mukhanov, Feldman, \&
  Brandenberger}]{mukhanov/feldman/brandenberger:1992}
Mukhanov, V.~F., Feldman, H.~A., \& Brandenberger, R.~H. 1992, Phys. Rept.,
  215, 203

\bibitem[{{Mukherjee} \& {Wang}(2003{\natexlab{a}})}]{mukherjee/wang:2003a}
{Mukherjee}, P. \& {Wang}, Y. 2003{\natexlab{a}}, 1562, \apj, submitted
  (astro-ph/0301562)

\bibitem[{{Mukherjee} \& {Wang}(2003{\natexlab{b}})}]{mukherjee/wang:2003b}
---. 2003{\natexlab{b}}, 1058, \apj, submitted (astro-ph/0301058)

\bibitem[{{Page} et~al.(2003)}]{page/etal:2003c}
{Page}, L. et~al. 2003, \apj, submitted

\bibitem[{Parker(1969)}]{parker:1969}
Parker, L. 1969, Phys. Rev., 183, 1057

\bibitem[{{Pearson} et~al.(2002)}]{pearson/etal:2002}
{Pearson}, T.~J., et~al. 2002, \apj, submitted (astro-ph/0205388)

\bibitem[{{Peebles} \& {Yu}(1970)}]{peebles/yu:1970}
{Peebles}, P.~J.~E. \& {Yu}, J.~T. 1970, \apj, 162, 815

\bibitem[{Pen et~al.(1994)Pen, Spergel, \& Turok}]{pen/spergel/turok:1994}
Pen, U.-L., Spergel, D.~N., \& Turok, N. 1994, Phys. Rev., D49, 692

\bibitem[{{Percival} et~al.(2001)}]{percival/etal:2001}
{Percival}, W.~J., et~al. 2001, \mnras, 327, 1297

\bibitem[{{Pierpaoli} et~al.(1999){Pierpaoli}, {Garcia-Bellido}, \&
  {Borgani}}]{pierpaoli/garcia-bellido/borgani:1999}
{Pierpaoli}, E., {Garcia-Bellido}, J., \& {Borgani}, S. 1999, Journal of High
  Energy Physics, 10, 15

\bibitem[{Polarski \& Starobinsky(1995)}]{polarski/starobinsky:1995}
Polarski, D. \& Starobinsky, A.~A. 1995, Phys. Lett., B356, 196

\bibitem[{Sasaki \& Stewart(1996)}]{sasaki/stewart:1996}
Sasaki, M. \& Stewart, E.~D. 1996, Prog. Theor. Phys., 95, 71

\bibitem[{{Sato}(1981)}]{sato:1981}
{Sato}, K. 1981, \mnras, 195, 467

\bibitem[{Seljak et~al.(1997)Seljak, Pen, \& Turok}]{seljak/pen/turok:1997}
Seljak, U., Pen, U.-L., \& Turok, N. 1997, Phys. Rev. Lett., 79, 1615

\bibitem[{{Seljak} \& {Zaldarriaga}(1996)}]{seljak/zaldarriaga:1996}
{Seljak}, U. \& {Zaldarriaga}, M. 1996, \apj, 469, 437

\bibitem[{Spergel \& Zaldarriaga(1997)}]{spergel/zaldarriaga:1997}
Spergel, D.~N. \& Zaldarriaga, M. 1997, \prl, 79, 2180

\bibitem[{{Spergel} et~al.(2003)}]{spergel/etal:2003}
{Spergel}, D.~N. et~al. 2003, \apj, submitted

\bibitem[{Starobinsky(1982)}]{starobinsky:1982}
Starobinsky, A.~A. 1982, Phys. Lett., B117, 175

\bibitem[{Stewart \& Lyth(1993)}]{stewart/lyth:1993}
Stewart, E.~D. \& Lyth, D.~H. 1993, Phys. Lett., B302, 171

\bibitem[{{Tsujikawa} et~al.(2002){Tsujikawa}, {Brandenberger}, \&
  {Finelli}}]{tsujikawa/brandenberger/finelli:2002}
{Tsujikawa}, S., {Brandenberger}, R., \& {Finelli}, F. 2002, \prd, 66, 83513

\bibitem[{{Turok}(1996{\natexlab{a}})}]{turok:1996b}
{Turok}, N. 1996{\natexlab{a}}, Phys. Rev. Lett., 77, 4138

\bibitem[{{Turok}(1996{\natexlab{b}})}]{turok:1996}
---. 1996{\natexlab{b}}, \apjl, 473, L5

\bibitem[{Turok et~al.(1998)Turok, Pen, \& Seljak}]{turok/pen/seljak:1998}
Turok, N., Pen, U.-L., \& Seljak, U. 1998, Phys. Rev., D58, 023506

\bibitem[{{Verde} et~al.(2003)}]{verde/etal:2003}
{Verde}, L. et~al. 2003, \apj, submitted

\bibitem[{Wands et~al.(2002)Wands, Bartolo, Matarrese, \&
  Riotto}]{wands/etal:2002}
Wands, D., Bartolo, N., Matarrese, S., \& Riotto, A. 2002, Phys. Rev., D66,
  043520

\bibitem[{{Wang} et~al.(1999){Wang}, {Spergel}, \&
  {Strauss}}]{wang/spergel/strauss:1999}
{Wang}, Y., {Spergel}, D.~N., \& {Strauss}, M.~A. 1999, \apj, 510, 20

\bibitem[{Zaldarriaga \& Harari(1995)}]{zaldarriaga/harari:1995}
Zaldarriaga, M. \& Harari, D.~D. 1995, Phys. Rev., D52, 3276

\end{thebibliography}

\clearpage


\end{document}